\documentclass[notitlepage,nofootinbib,preprintnumbers,aps,prd,superscriptaddress]{revtex4-1}
\usepackage{amsmath,amssymb,natbib,bm,color}
\usepackage{tikz}
\usetikzlibrary{positioning,decorations.pathmorphing,decorations.markings,shapes.symbols}
\usepackage{appendix}
\usepackage{comment}
\usepackage{url}
\usepackage[normalem]{ulem}

\newcommand{\mrm}[1]{_{\rm #1}}
\newcommand{\DNeff}{\Delta N_{\rm eff}}

\begin{document}
\preprint{\hbox{CERN-TH-2021-049}}

\title{Precision Calculation of Dark Radiation from Spinning Primordial Black Holes and Early Matter Dominated Eras}
\author{Alexandre Arbey}
\affiliation{Univ. Lyon, Univ. Claude Bernard Lyon 1, CNRS/IN2P3, IP2I Lyon, UMR 5822, F-69622, Villeurbanne, France}
\affiliation{Theoretical Physics Department, CERN, CH-1211 Geneva 23, Switzerland}
\affiliation{Institut Universitaire de France (IUF), 103 boulevard Saint-Michel, 75005 Paris, France}
\author{J\'er\'emy Auffinger}
\affiliation{Univ. Lyon, Univ. Claude Bernard Lyon 1, CNRS/IN2P3, IP2I Lyon, UMR 5822, F-69622, Villeurbanne, France}
\author{Pearl Sandick}
\affiliation{Department of Physics and Astronomy, University of Utah, Salt Lake City, UT 84112, USA}
\author{Barmak Shams Es Haghi}
\affiliation{Department of Physics and Astronomy, University of Utah, Salt Lake City, UT 84112, USA}
\author{Kuver Sinha}
\affiliation{Department of Physics and Astronomy, University of Oklahoma, Norman, OK 73019, USA}

\begin{abstract}

We present precision calculations of dark radiation in the form of gravitons coming from Hawking evaporation of spinning primordial black holes (PBHs) in the early Universe. Our calculation incorporates a careful treatment of extended spin distributions of a population of PBHs, the PBH reheating temperature, and the number of relativistic degrees of freedom. 
We compare our precision results with those existing in the literature, and show constraints on PBHs from current bounds on dark radiation from BBN and the CMB, as well as the projected sensitivity of CMB Stage 4 experiments.
As an application, we consider the case of PBHs formed during an early matter-dominated era (EMDE). We calculate graviton production from various PBH spin distributions pertinent to EMDEs, and find that PBHs in the entire mass range up to $10^9\,$g will be constrained by measurements from CMB Stage 4 experiments, assuming PBHs come to dominate the Universe prior to Hawking evaporation. 
We also find that for PBHs with monochromatic spins $a^*>0.81$, all PBH masses in the range
$10^{-1}\,{\rm g} < M_{\rm BH} <10^9\,$g
will be probed by CMB Stage 4 experiments.
\end{abstract}

\maketitle

\section{Introduction}

Black hole evaporation via the emission of Hawking radiation is a well established phenomenon~\cite{Hawking:1974rv,Hawking:1974sw}, with recent work towards precisely characterizing the Hawking radiation yields of relevant particles and the time evolution of the population of black holes (\textit{e.g.}~\cite{Arbey:2019mbc}).  Primordial black holes (PBHs) are of particular interest in that their possible mass range spans many orders of magnitude and they could be relevant to the questions of dark matter and cosmological chronology, and their existence can affect observable quantities that can be probed with current (and future) cosmological experiments.  Here, we undertake a precision study of Hawking evaporation of PBHs prior to Big Bang Nucleosynthesis (BBN), with particular attention to PBH spin and spin distributions, the PBH reheating temperature, and the evolution of the number of relativistic degrees of freedom, and compare our results to the current sensitivities from the cosmic microwave background (CMB) and BBN, as well as future CMB Stage 4 experiments. 

PBHs may have formed in the early Universe from the collapse of primordial density inhomogeneities originating from quantum fluctuations prior to inflation or from topological defects such as cosmic strings or domain walls. Bubble collisions during a first-order phase transition can also trigger PBH formation. For a recent review of PBH formation mechanisms, we refer to~\cite{Carr:2020gox} and references therein. 

The spin of the resulting PBH population depends on the equation of state (as does the mass distribution). PBHs formed during radiation domination are believed to have negligible spin~\cite{DeLuca:2019buf}. On the other hand, PBHs formed during an early matter-dominated era (EMDE) \cite{Kane:2015jia,  Georg:2019jld, Georg:2017mqk, Georg:2016yxa} could have sizeable to near-extremal spin~\cite{Allahverdi:2020bys, Harada:2016mhb, Harada:2017fjm}.  PBHs can also  accumulate some spin either through early accretion processes~\cite{DeLuca:2019buf} or through hierarchical mergers~\cite{Fishbach:2017dwv}. 
In the last two decades, constraints have been placed on a wide range of PBH masses, assuming Schwarzschild (non-rotating) PBHs with monochromatic mass spectra\footnote{The distribution of PBHs can also have an extended mass function, for example if the power spectrum of primordial inhomogeneities embeds a wide peak around some spatial scale~\cite{Carr:2016drx,Carr:2017jsz}.  
Extended mass functions of spinning PBHs have not yet been thoroughly studied. We leave this for future work.} (for a review see \textit{e.g.}~\cite{Carr:2020gox}).
Using a combination of numerical and analytical results for Hawking radiation, recent studies have started to complete the constraints on PBHs with non-zero spin \cite{Dong:2015yjs,Arbey:2019vqx,Dasgupta:2019cae,Laha:2020vhg,Hooper:2020evu,Ray:2021mxu,Masina:2021zpu}.

Here we study the production of dark radiation in the form of gravitons coming from Hawking evaporation of populations of spinning PBHs prior to BBN.  We compute the primary and secondary spectra of Standard Model (SM) particles and gravitons for realistic spin distributions of PBHs from an EMDE~\cite{Harada:2016mhb} as well as a hierarchical merger history~\cite{Fishbach:2017dwv}. 
Our calculations are performed with the public code \texttt{BlackHawk}~\cite{Arbey:2019mbc}, developed by a subset of the current authors\footnote{We have implemented the possibility of adding a particle to the SM, \textit{e.g.}~the massless spin 2 graviton or general dark sector particles of spin 0, 1, 2 or $\frac{1}{2}$  in \texttt{BlackHawk}, although in this study we focus only on massless spin 2 graviton emission. Additional dark sector particles have not yet been implemented in the public version of the code. To our knowledge, this is the first precision calculation of Hawking radiation with with non-trivial PBH spin distributions using \texttt{BlackHawk}.}. 
The evolution of a given distribution of PBHs and the associated time-dependent spectrum of emitted gravitons are computed, allowing a straightforward determination of the total energy emitted in the form of dark radiation.  This affects the number of relativistic species, with the result characterized as the deviation from the SM expectation of the effective number of neutrino species, $\DNeff$.  We compute 
$\DNeff$, and compare it to existing results in the literature and interpret it in the context of current limits on $\DNeff$ from BBN and CMB measurements. In particular, we carefully calculate BBN constraints on the dark radiation density using \texttt{AlterBBN}~\cite{Arbey:2011nf,Arbey:2018zfh}.

The main application of our results is the calculation of $\DNeff$ from PBHs that were formed during an EMDE and subsequently came to dominate the Universe prior to Hawking evaporation. EMDEs are  highly motivated due to the ubiquity of moduli in string theory and have been extensively studied in recent years in the context of dark matter \cite{Dutta:2009uf, Allahverdi:2012gk, Acharya:2009zt, Acharya:2008bk, Erickcek:2015bda, Delos:2019dyh} and baryogenesis \cite{Allahverdi:2010im}. Detailed studies of PBHs formed during an EMDE have been performed by \cite{Georg:2019jld, Georg:2017mqk, Georg:2016yxa}, with a focus on long-lived PBHs existing in the current Universe, and their interplay with dark matter physics. PBHs that evaporated before BBN are harder to constrain\footnote{The authors of~\cite{Masina:2020xhk,Auffinger:2020afu,Masina:2021zpu,Gondolo:2020uqv} considered PBHs that evaporated before BBN and gave rise to non-thermal dark matter.}. The authors of \cite{Matsubara:2019qzv, Kokubu:2018fxy, Harada:2016mhb, Harada:2017fjm} have initiated much progress in this direction; of particular relevance for our work are the formation rate \cite{Harada:2016mhb} and spin distribution \cite{Harada:2017fjm} of PBHs formed during an EMDE.  Following the spin  distributions used in~\cite{Harada:2017fjm} as benchmark examples, we find that PBHs formed during an EMDE with a spin distribution due to the first-order effect are constrained by current CMB bounds on $\DNeff$ in the mass range $10^{8}-10^{9}\,$g; they are completely constrained in the mass range $10^{-1}-10^{9}\,$g by projections of CMB Stage 4 experiments. PBHs that formed during an EMDE with spin distribution due to the second-order effect, on the other hand, are not constrained by current BBN or CMB bounds on $\DNeff$; they too would, however, be completely constrained in the mass range $10^{-1}-10^{9}\,$g by CMB Stage 4 projections (Fig.~\ref{fig:2_b}).

The fact that PBHs formed during an EMDE that evaporate before BBN will be completely probed by $\DNeff$ measurements from CMB Stage 4 experiments is the main result of our work. Physically, this happens because PBHs formed during an EMDE are endowed with significant spin, which enhances their production of gravitons during evaporation. It should be noted that the $\DNeff$ constraints are only relevant if the PBHs come to dominate the Universe. Generally, this is quite restrictive on the sector that causes the EMDE. We consider a gravitationally coupled modulus that causes the EMDE and obtain conditions on the decay width (and hence the modulus mass) such that this condition holds. In terms of the modulus sector, our result is that for a variety of PBH spin distributions and fractions $\beta$ of the total energy density of the Universe that is constituted by PBHs at formation time during an EMDE, moduli with masses larger than $\sim 10^8$ GeV will be constrained by CMB Stage 4 experiments (Fig. \ref{fig:order2fig}).

We also consider the case of a spin distribution due to inspirals of PBHs under a heirarchical merger history, obtaining, for the first time, precision predictions for $\DNeff$ in this scenario, which will be probed by CMB Stage 4 experiment.
Finally, we go on to apply our results to the case of PBHs with extremal spins regardless of origin, and find that PBHs with spin $a^* \gtrsim 0.99$ and mass $M\mrm{BH} \gtrsim 10^8\,$g are excluded by  CMB stringent constraints (TT,TE,EE+low E) while those with even higher spin $a^* \gtrsim 0.999$ are constrained by the CMB conservative constraints (TT+low E), but only for masses $M\mrm{BH} \gtrsim 2\times 10^8\,$g. We further determine that the limiting value of the  PBH spin that will be constrained by CMB Stage 4 experiment for \textit{all} PBH masses up to $10^9\,$g is $a^*\mrm{min,\,all} \simeq 0.81\,$.

Our paper is organized as follows. In Section~\ref{sec:PBH}, we give an overview of the formation and evaporation of Kerr PBHs.  In Section~\ref{sec:BlackHawk} we outline the precision calculation of the effective number of neutrino species, $\DNeff$, from PBH evaporation, addressing spin distributions and the reheating temperature in Subsections~\ref{sec:spindist} and~\ref{sec:reheat}, respectively.  We present the bulk of our results in Section~\ref{sec:results}. In Section~\ref{sec:comparison}, we compare precision results for benchmark spins versus spin distributions, including the effects of the reheating temperature and a precision accounting of the effective degrees of freedom. In Section~\ref{emdeetc},  we explicitly focus on spin distributions relevant for an EMDE. The effect on BBN is discussed in Section~\ref{sec:BBN}, and our conclusions are given in Section~\ref{sec:conclusion}.  Finally, we include three appendices, where we discuss the details of PBH formation and evaporation during an EMDE, PBH spin distributions from an EMDE, and PBH spin distributions from inspirals.

\section{Kerr primordial black holes: formation and evaporation}
\label{sec:PBH}
Hawking has demonstrated that black holes evaporate~\cite{Hawking:1974rv, Hawking:1974sw} by emitting quasi-thermal radiation with a temperature 
\begin{equation}
    T_{\rm S} = \dfrac{1}{8\pi M\mrm{BH}}\,,
\end{equation}
for the Schwarzschild solution,
and
\begin{equation}
    T_{\rm K} = \dfrac{1}{2\pi}\left( \dfrac{r_+ - M\mrm{BH}}{r_+^2 + a^{*2} M\mrm{BH}^2} \right),
\end{equation}
for the Kerr solution\footnote{In these equations and in the rest of the paper, we use the natural system of units $G = \hbar = k\mrm{B} = c = 1$.}~\cite{Dong:2015yjs}.  For a black hole with angular momentum $L$ and mass $M\mrm{BH}$, the dimensionless black hole angular momentum, or spin, is 
\begin{equation}
    a^* \equiv L/M\mrm{BH}^2\,, 
\end{equation}
and the exterior horizon is given by
\begin{equation}
    r_+ \equiv M\mrm{BH}(1 + \sqrt{1 - a^{*2}})\,.
\end{equation} 
The rate of emission of one degree of freedom of a particle $i$ per unit time and energy is given by
\begin{equation}
    \dfrac{{ d}^2 N_i}{{ d}t{ d}E} = \dfrac{1}{2\pi}\dfrac{\Gamma_{s_i}^{l,m}}{e^{E^\prime/T_{\rm K}} - (-1)^{2s_i}}\,,\label{eq:master_HR}
\end{equation}
where $s_i$ is the particle spin, $E^\prime \equiv E - m\Omega = E - ma^*/2r_+$ is the particle energy corrected for horizon rotation and $m$ is the projection of the particle's angular momentum $l$. The quantity $\Gamma_{s_i}^{lm}$, the so-called greybody factor, describes the probability that a Hawking radiated particle escapes the gravitational well of the black hole to spatial infinity. In general, it depends on the particle angular momentum numbers $(l,m)$, energy $E$, and spin $s_i$, and on the black hole mass and spin: $\Gamma_{s_i}^{l,m}(E,M\mrm{BH},a^*)$. It should also depend on the particle rest mass $\mu_i$ but as an approximation we will consider (as in \texttt{BlackHawk}) that the particle rest mass acts as a cut-off at $E<\mu_i$ in the particle emission spectrum.

Due to this continuous emission of all degrees of freedom (SM and beyond), black holes lose mass and angular momentum\footnote{Angular momentum is lost because on average, the coupling between the black hole and the particle's angular momentum favors the emission of aligned spin modes.}. This can be described using the Page factors $f(M\mrm{BH},a^*)$ and $g(M\mrm{BH},a^*)$ \cite{Page:1976df,Dong:2015yjs} which are the result of integration over all degrees of freedom (dof) that a black hole with mass $M\mrm{BH}$ can emit:
\begin{align}
    &f(M\mrm{BH},a^*) \equiv -M^2 \dfrac{{ d} M\mrm{BH}}{{ d} t} = M\mrm{BH}^2\int_{0}^{+\infty} \sum_i\sum_{\rm dof} \dfrac{E}{2\pi}\dfrac{\Gamma_{s_i}^{l,m}(E,M\mrm{BH},a^*)}{e^{E^\prime/T_{\rm K}} -  (-1)^{2s_i}} { d} E\,, \label{eq:Pagef} \\
	&g(M\mrm{BH},a^*) \equiv -\dfrac{M\mrm{BH}}{a^*} \dfrac{{ d} L}{{ d} t} = \dfrac{M\mrm{BH}}{a^*}\int_{0}^{+\infty} \sum_i\sum_{\rm dof}\dfrac{m}{2\pi} \dfrac{\Gamma_{s_i}^{l,m}(E,M\mrm{BH},a^*)}{e^{E^\prime/T_{\rm K}} -(-1)^{2s_i}} { d} E\,, \label{eq:Pageg}
\end{align}
where the sum over the degrees of freedom accounts for angular momentum degrees of freedom as well as polarization/color multiplicity of particle $i$. Using the definitions of $f$ and $g$, it is straightforward to write differential equations for the evolution of the black hole mass and spin,
\begin{align}
    &\dfrac{d M\mrm{BH}}{d t} = -\dfrac{f(M\mrm{BH},a^*)}{M\mrm{BH}^2}\,, \\
    &\dfrac{d a^*}{d t} = \dfrac{a^*\left[2f(M\mrm{BH},a^*) - g(M\mrm{BH},a^*)\right]}{M\mrm{BH}^3}\,.
\end{align}

Any degree of freedom additional to the SM would be Hawking emitted as this process is purely gravitational. This would increase the Page factors, Eqs.~\eqref{eq:Pagef} and \eqref{eq:Pageg}, and hasten the black hole disappearance. In the case we study here, \textit{i.e.}~additional emission of spin 2 massless gravitons, the number of added degrees of freedom (2) compared to the SM is very small, and thus the effect on the Page factors is negligible, so the lifetime of PBHs remains essentially unchanged. Nonetheless this effect is taken into account in \texttt{BlackHawk}. 

Recent studies have tried to constrain the fraction of ultra-light PBHs with masses  $ 10^{-5} \,\rm{g}\lesssim M\mrm{BH}\lesssim 10^{9}\,$g by considering that they emit dark sector particles before BBN. This mass range is unconstrained by current cosmological observations (though may be probed by future gravitational wave experiments~\cite{Papanikolaou:2020qtd}). This scenario would therefore be an elegant way of providing the (warm) dark matter content of the Universe while evading PBH constraints~\cite{Fujita:2014hha,Lennon:2017tqq,Baldes:2020nuv,Masina:2020xhk,Gondolo:2020uqv,Auffinger:2020afu}. If sufficiently light, this energetic dark sector can provide dark radiation that can measurably affect cosmology, which we will review in the next sections.

\section{Precision $\Delta N_\text{eff}$ Calculations}
\label{sec:BlackHawk}

Hawking evaporation of PBHs in the early Universe creates SM particles along with other particles that are either decoupled or feebly interacting with the SM. In this Section, we outline the steps for calculating $\DNeff=N_{\rm eff} - 3.046$, where $N_{\rm eff}$ is the total number of relativistic degrees of freedom and 3.046 is the SM expectation, from PBH evaporation.
The precision calculations involve two steps: taking into account the distribution of PBH spins and carefully defining the reheating temperature. We also use a precise expression for the number of accessible degrees of freedom.

First we review the standard calculation of $\Delta N_\text{eff}$.  Using conservation of entropy during the expansion of the Universe, one can track the evolution of the energy density of dark radiation from reheating to matter-radiation equality.
For a population of PBHs with lifetime $\tau$, the age of the Universe at formation is small relative to $\tau$ such that the evaporation 
time is $t_\text{eva}\simeq\tau$.
Assuming instantaneous thermalization of SM particles at the end of PBH evaporation, the reheating temperature, $T_\text{RH}$, can be obtained as
\begin{equation}
\rho_\text{PBH}(\tau)-\rho_\text{DR}(\tau)=\rho_\text{SM}(\tau)\equiv \frac{\pi^2}{30}g_*(T_\text{RH})T_\text{RH}^4\,, \label{eq:rad_dens}
\end{equation}
where $\rho_\text{PBH}(\tau)$ is the energy density of PBHs at the time of evaporation, $\rho_\text{DR}$ ($\rho_\text{SM}$) is the amount of energy PBHs emit in the form of dark radiation (SM particles), and $g_*(T)$ denotes the total number of relativistic degrees of freedom at temperature $T$, given by
\begin{equation}
g_*(T)=\sum_B g_B\left(\frac{T_B}{T}\right)^4+\frac{7}{8}\sum_F g_F\left(\frac{T_F}{T}\right)^4.
\end{equation} 
Here the sum includes all bosonic ($B$) and fermionic ($F$) degrees of freedom with temperatures of $T_B$ and $T_F$, respectively. The density of PBHs at evaporation is related to the density of PBHs at formation, usually expressed in terms of the fraction of the energy density of the Universe that collapsed into PBHs at PBH formation time, which is denoted by $\beta$. In this work, we assume that $\beta$ is sufficiently large such that the energy density of PBHs exceeds that of radiation at some time before evaporation. A discussion of such a scenario is given in Appendix~\ref{emdemodel} in the case of modulus decay. With this hypothesis, the density of PBHs at evaporation is fixed by the fact that SM radiation produced by PBH  Hawking evaporation constitutes the main component of SM radiation at reheating. Thus, tracing the redshifted temperature of the CMB today back to reheating (from today back to the matter-radiation equality time with $a(t)\sim t^{2/3}$ and then to the reheating time, $t_{\rm{RH}}\simeq\tau$, with  $a(t)\sim t^{1/2}$), we obtain the value of $T\mrm{RH}$. The values we obtain for $\DNeff$ in this study should be considered as \textit{upper limits} in the case of full PBH domination prior to evaporation. The constraints are generally weakened but must be recalculated if PBHs do not dominate the energy density of the Universe before evaporation.

The energy density of SM radiation (all relativistic particles) is therefore diluted as
\begin{equation}
    \frac{\rho_\text{R}(t_\text{EQ})}{\rho_\text{R}(t_\text{RH})}=\left(\frac{a_\text{RH}}{a_\text{EQ}}\right)^4
    \left(\frac{g_{*}(T_\text{EQ})}{g_{*}(T_\text{RH})}\right)\left(\frac{g_{*,S}(T_\text{RH})}{g_{*,S}(T_\text{EQ})}\right)^{4/3},
\end{equation}
where $a_\text{RH(EQ)}$ is the scale factor at reheating (matter-radiation equality), 
and $g_{*,S}(T)$ counts the number of relativistic degrees of freedom contributing to the entropy, given by
\begin{equation}
g_{*,S}(T)=\sum_B g_B\left(\frac{T_B}{T}\right)^3+\frac{7}{8}\sum_F g_F\left(\frac{T_F}{T}\right)^3.
\end{equation}
Similarly, the energy density of dark radiation, $\rho_\text{DR}$, also dilutes as
\begin{equation}
    \frac{\rho_\text{DR}(t_\text{EQ})}{\rho_\text{DR}(t_\text{RH})}=\left(\frac{a_\text{RH}}{a_\text{EQ}}\right)^4.
\end{equation}
Therefore, the ratio of the energy density of dark radiation to the SM radiation energy density at matter-radiation equality becomes
\begin{equation}
     \frac{\rho_\text{DR}(t_\text{EQ})}{\rho_\text{R}(t_\text{EQ})}=\frac{\rho_\text{DR}(t_\text{RH})}{\rho_\text{R}(t_\text{RH})}
     \left(\frac{g_{*}(T_\text{RH})}{g_{*}(T_\text{EQ})}\right)\left(\frac{g_{*,S}(T_\text{EQ})}{g_{*,S}(T_\text{RH})}\right)^{4/3} , \label{eq:gstar}
\end{equation}
which determines the effective number of neutrino species as \cite{Hooper:2019gtx}
\begin{equation}
    \Delta N_\text{eff}=\frac{\rho_\text{DR}(t_\text{EQ})}{\rho_\text{R}(t_\text{EQ})}\left[N_\nu+\frac{8}{7}\left(\frac{11}{4}\right)^{4/3}\right]. \label{eq:DNeff}
\end{equation}


\subsection{Extended PBH spin distributions}
\label{sec:spindist}

A monochromatic distribution of non-rotating PBHs is only a convenient approximation to the more realistic extended mass distribution of rotating PBHs generated by detailed models of PBH formation, accretion, and mergers. For the purpose of this study, we focus on single-mass, rotating PBHs with a spin number distribution $n(a^*)$ normalized to unity,
\begin{equation}
    \int_{0}^1 \dfrac{d n}{d a^*}\,d a^* = 1\,.\label{eq:normalization}
\end{equation}
Note that the assumption of a monochromatic mass distribution for PBHs is justified if the PBH production occurs at a precise time, leading to a very narrowly peaked mass distribution.
The total energy that has been emitted in the form of dark radiation by the reheating time $t_{\rm RH}$ can be expressed as a ratio over the SM emission, \textit{i.e.}~the ratio of the energy densities after evaporation is complete,
\begin{equation}
    f_{\rm DR} \equiv \dfrac{\rho_{\rm DR}(t_{\rm RH})}{\rho_{\rm SM}(t_{\rm RH})} = \dfrac{\rho\mrm{DR}(t_{\rm RH})}{\rho\mrm{BH}(t_{\rm RH}) - \rho\mrm{DR}(t_{\rm RH})}\,,\label{eq:ratio}
\end{equation}
where $\rho\mrm{DR/SM}(t_{\rm RH})$ is the total emission integrated over the history of the Universe prior to reheating,
\begin{equation}
    \rho\mrm{DR/SM}(t_{\rm RH}) = \int_{0}^{1} d a^*\, \dfrac{d n}{d a^*}\int_{0}^{t_{\rm RH}} d t  \int_{0}^{+\infty} d E\, E\,\dfrac{d^2 N_{\rm DR/SM}}{d t d E}(M,a^*)\,,
\end{equation}
and
\begin{equation}
    \dfrac{d^2 N_{\rm SM}}{d t d E} \equiv \sum_{i\,\in\,\text{SM}} \dfrac{d^2 N_i}{d t d E}\,.
\end{equation}
The emission rates for individual species, $d^2N_i/d t d E$, come from Eq.~\eqref{eq:master_HR}.
We stress that the ratio \eqref{eq:ratio} takes into account the fact that for high DR emission, which occurs for highly spinning black holes, the approximation $\rho\mrm{BH}\simeq\rho\mrm{R}$ used in~\cite{Hooper:2020evu,Masina:2021zpu} no longer holds. This could be one of the reasons our results differ from those of~\cite{Masina:2021zpu} for high PBH spin. We recall that $\rho\mrm{SM} = \rho\mrm{R}$ at time $t_{\rm RH}$ (which occurs before matter-radiation equality), which allows to use the ratio \eqref{eq:ratio} in Eq.~\eqref{eq:gstar} to determine $\DNeff$.  Furthermore, we note that the normalization of the density of PBHs $\rho_{\rm BH}$ is irrelevant to the computation of $\DNeff$, since it cancels out of the ratio $f_{\rm DR}$ in Eq.~\eqref{eq:ratio}.

For the purposes of this study, we have implemented in \texttt{BlackHawk} the possibility of including additional particles beyond those in the SM, \textit{e.g.}~the massless spin 2 graviton or general dark sector particles of spin 0, 1, 2 or $\frac{1}{2}$ (although here we focus only on gravitons). We compute the evolution of a given  distribution of PBHs and the associated time-dependent spectrum for this additional particle\footnote{Adding a particle to the \texttt{BlackHawk} spectra has already been done for warm dark matter calculations~\cite{Auffinger:2020afu}.}. It is then straightforward to integrate over this spectrum to obtain the total energy emitted in the form of SM particles and additional dark radiation and hence the ratio $f_{\rm DR}$ in Eq.~\eqref{eq:ratio}.

The main effect of a spin distribution, relative to monochromatic spin, is to modify the rate of emission of dark radiation, and thus its ratio to SM radiation, as in Eq.~\eqref{eq:ratio}. Indeed, it is well known that spinning black holes emit more high spin particles ($s_i = 1$ or $s_i = 2$) than non-spinning black holes. As we consider the emission of spin 2 massless gravitons, this effect can be quite sizeable, with the emission being enhanced by a factor of up to $\sim 10^4$~\cite{Page:1976ki}. The effect of this enhancement on the ratio~\eqref{eq:ratio} is somewhat less dramatic, since the emission of spin 0, 1, and $\frac{1}{2}$ SM particles also increases. Still, taking into account extended spin distributions of PBHs with significant high-spin component enhances $f\mrm{DR}$ and hence  $\DNeff$, leading to more stringent constraints than one would find for simple single-spin distributions.

For the greybody factors $\Gamma_{s_i}^{l,m}$ of Eq.~\eqref{eq:master_HR}, we use tabulated values computed by solving the Teukolsky equations for spinning black holes and all particle spins 0, 1, 2, $\frac{1}{2}$ (for more details, see the \texttt{BlackHawk} manual~\cite{Arbey:2019mbc}). Note that we deviate from the procedures in~\cite{Dong:2015yjs} by solving Schr\"odinger-like wave equations with short-range potentials to find $\Gamma_s^{l,m}$, which also ensures robust numerical stability of the result. For one particle $i$, the emission is summed over angular momenta $l = s_i, s_i+1, ...$ and their projections $m = -l, ..., l$ recursively until some asymptotic value is reached (in practice, we do not need to go beyond $l \sim 30$). We pay particular attention in \texttt{BlackHawk} to distributions of PBHs; this is a great improvement over previous studies of
the abundance of PBHs (see \textit{e.g.}~\cite{Arbey:2019vqx,Laha:2019ssq,Dasgupta:2019cae,Chan:2020zry,Luo:2020dlg}). To our knowledge, this is the first precision calculation of Hawking radiation from a population of PBHs with any non-trivial spin distribution.

\subsection{Reheating temperature and degrees of freedom}
\label{sec:reheat}

When an extended spin distribution of PBHs is employed rather than a monochromatic spin distribution, there is some subtlety in defining the reheating temperature. As spinning black holes emit more radiation than non-spinning ones, with a continuous increase in the emission as $a^*$ increases, they evaporate faster. 
Although initial nonzero spin has a small effect on black hole lifetime (somewhat less than $60\%$ diminution for extremal spin~\cite{Page:1976ki,Arbey:2019jmj}), it does influence the way one defines the reheating time. For PBHs with lifetime $\tau$ ({\it e.g.} for PBHs with monochromatic mass and spin distributions), assuming an instantaneous reheating in Eq.~\eqref{eq:rad_dens} is justified by the fact that PBHs emit most of their Hawking radiation during a period of time that is negligibly small relative to their lifetime. However, since black holes with higher spin evaporate faster than black holes with lower spin, a distribution in initial spins causes a spread of the evaporation times and a non-instantaneous reheating scenario. 

For simplicity, here we consider two possibilities for the definition of the reheating time: 
\begin{enumerate}
\item the reheating time corresponds to the time at which the last PBHs (with the lowest spins) evaporate; and 
\item the reheating time corresponds to the average PBH lifetime, weighted by the spin distribution,
\begin{equation}
    \langle \tau \rangle  \equiv \int_0^1 \tau(M,a^*)\dfrac{d n}{d a^*}\,{ d}a^*\,. \label{eq:average_tau}
\end{equation}
\end{enumerate}
We believe that the second option is more physically realistic, as the averaged lifetime corresponds roughly to the peak of the emission of the Hawking radiation. 
We discuss both options in Section~\ref{sec:results}, where we present our results.

Finally, in order to obtain the ratio \eqref{eq:gstar}, it is necessary to specify the quantities $g_{*}(T)$ and $g_{*,S}(T)$. We stress that precise determination of these numbers of degrees of freedom are model-dependent, especially for the region of temperatures close to the QCD phase transition.  Here, that corresponds to $M\mrm{BH}\sim 7\times 10^8\,$g ($T\sim 100\,$MeV). Refs.~\cite{Masina:2020xhk,Masina:2021zpu} use step functions which give results qualitatively similar to ours, while the model used in~\cite{Hooper:2020evu} is not made explicit and shows a significantly different behaviour. In this work, we use the tabulated values of $g_{*}(T)$ and $g_{*,S}(T)$ available with the public code \texttt{SuperIso Relic}\footnote{The code can be obtained at \url{http://superiso.in2p3.fr/relic/}}~\cite{Arbey:2009gu,Arbey:2018msw}.

\section{Precision results for $\DNeff$}
\label{sec:results}

Here, we present precision results for $\DNeff$ with improvements to the calculation as described above.  In Subsection~\ref{sec:comparison}, we explore the effect of each of the three precision elements we have included here; spin distributions, reheating temperature, and degrees of freedom. In Subsection~\ref{sec:spin_dist_MD} we present, for the first time, explicit predictions for $\DNeff$ from PBH spin distributions expected from an EMDE.  

In all cases, we compare our results for $\DNeff$ to current experimental limits and projected sensitivities of future experiments.  We present three relevant CMB constraints/sensitivities: two are taken from the Planck Collaboration~\cite{Aghanim:2018eyx} and are denoted in the plots as CMB$^1$ (TT+low E, conservative) and CMB$^2$ (TT,TE,EE+low E, more stringent). The third one is the sensitivity of the future CMB Stage 4 (CMB-S4) experiment, and represents an order of magnitude improvement over current limits (see details in~\cite{Abazajian:2016yjj,Baumann:2017gkg,Hanany:2019lle}). Where relevant, we also include the constraint on $\DNeff$ from BBN, as discussed in Section~\ref{sec:BBN}.

\subsection{Benchmark spin scenarios - exploring precision results}
\label{sec:comparison}

In this subsection we compute $\DNeff$, incorporating the precision calculations described above -- spin distributions, reheating temperature, and degrees of freedom -- for some benchmark PBH spin scenarios.  We compare the results for $\DNeff$ calculated with an extended spin distribution to those obtained from monochromatic spin distributions ({\it e.g.} the central/peak value of the extended distribution), as well as $\DNeff$ obtained with the two reheating temperature calculations.  Furthermore, we compare our results to previous calculations in the literature for $a^*=0$ and $a^*=0.99$ to demonstrate the full effects of the precision calculation. 

We first make a few comments about PBH masses in the low mass regime. In our calculations, we find that changing the PBH mass in the range $10^{-1}\,{\rm g} < M\mrm{BH} < 10^9\,{\rm g}$ has a very small effect on the ratio $\rho_{\rm DR}/\rho_{\rm R}$ (less than $1\%$ over the whole mass range).
This is because, for a given spin distribution, the main variation in $\DNeff$ as the PBH mass is varied comes from the different reheating times (and thus reheating temperatures). Below $M\mrm{BH} \lesssim 10^{5}\,$g, the reheating temperature is far above the mass of all the SM particles ($T\mrm{RH}\gg 10^2\,$GeV), so $g_{*}(T)$ and $g_{*,S}(T)$ have already reached their asymptotic values. Thus, $\DNeff$ values for $M\mrm{BH}\lesssim 10^{5}\,$g can be safely extrapolated from their value corresponding to the case of $M\mrm{BH} = 10^{5}\,$g. We note that our results also apply to the $M\mrm{BH} = 10^{-5}- 10^{-1}\,$g mass range for PBHs. This range is sometimes excluded from analyses due to model-dependent limits on the inflationary Hubble parameter~\cite{Akrami:2018odb,Masina:2021zpu}. Below, we present results only for $10^{5} \, \rm{g} \leq M\mrm{BH} \leq 10^{9}\,$g.

\begin{figure}
    \centering
    \includegraphics[scale = 1]{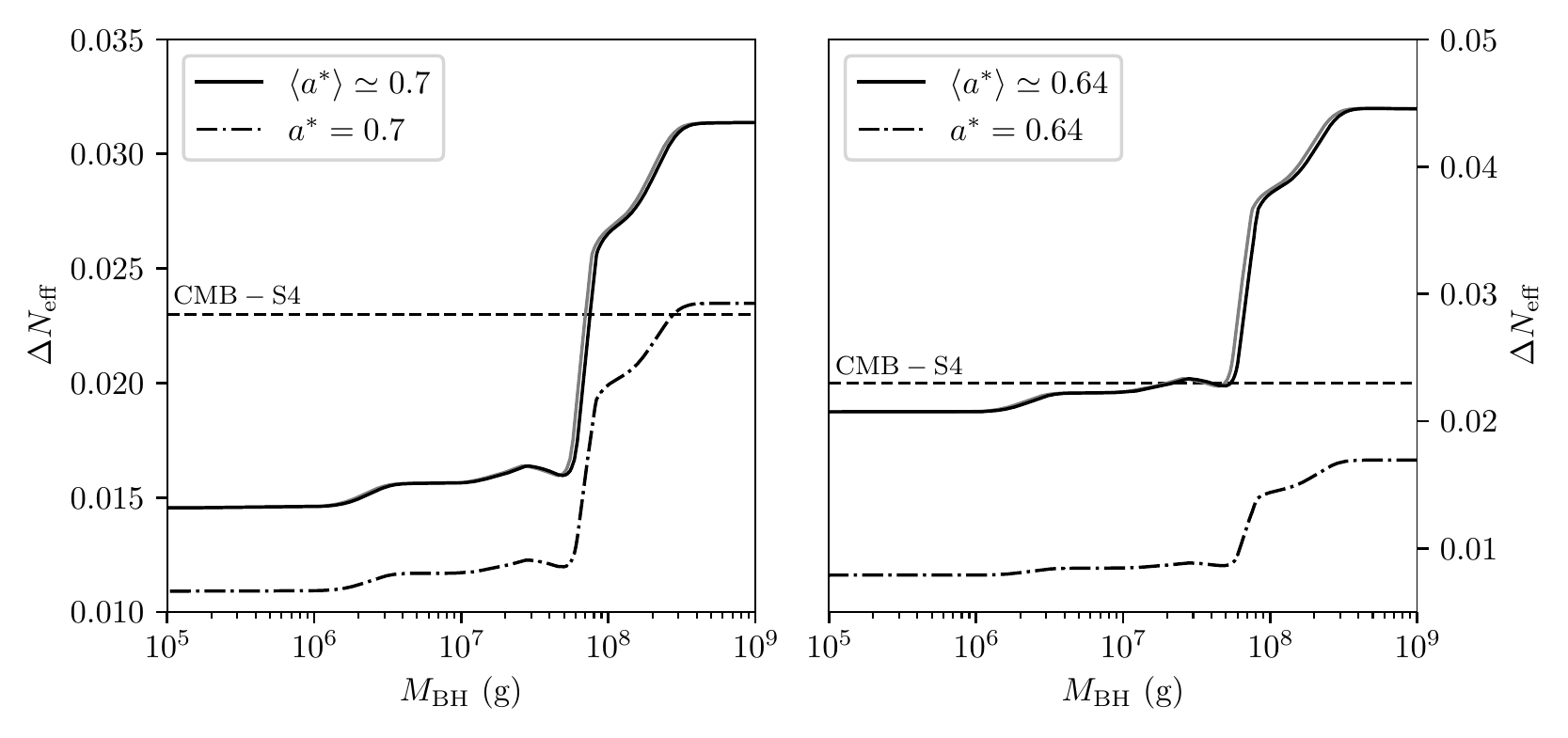}
    \caption{\textbf{Left:} Comparison of the results for $a^* = 0.7$ (dot-dashed line) and the full ``inspiral" distribution with $\left\langle a^* \right\rangle \simeq 0.7$ (solid lines). The relative difference in $\Delta N_\text{eff}$ between the solid and dashed curves is $\sim 25\%$. \textbf{Right:}  Comparison of the results for $a^* = 0.64$ (dot-dashed line) and a benchmark extended distribution from an EMDE with $\left\langle a^* \right\rangle \simeq 0.64$ (solid lines). The relative difference in $\Delta N_\text{eff}$ between the solid and dashed curves is $\sim 60\%$. The black (grey) curves correspond to instantaneous reheating at the weighted average value of the black hole lifetimes (last black hole evaporation). The prospective CMB-S4 constraint (horizontal dashed line) is extracted from~\cite{Hooper:2020evu}.}
    \label{fig:1_a_b}
\end{figure}

To show how the prediction for $\DNeff$ from an extended distribution of PBH spins compares to the monochromatic approximation, we present two benchmark extended spin distributions, along with the corresponding prediction assuming a monochromatic distribution.
We first consider the asymptotic spin distribution expected for multiple generation PBH inspirals~\cite{Fishbach:2017dwv} (see Appendix~\ref{sec:spin_dist_insp} for details). The average spin in this case is $\langle a^* \rangle \simeq 0.7$, so we compare the results for the full spin distribution to those for the monochromatic spin distribution with $a^* = 0.7$. As discussed above, we expect more gravitons to be emitted because there are higher spin PBHs in the extended distribution, relative to the monochromatic case. This is borne out in the results shown in the left panel of Fig.~\ref{fig:1_a_b}. 
We see that $\DNeff$ indeed does acquire greater values (by $\sim 25\%$) for the full distribution than for the monochromatic one. This discrepancy becomes critical for PBH masses above $7\times 10^7\,$g; in the case of the extended distribution, these PBHs will be probed by CMB-S4, while the average spin approximation leads to the conclusion that only PBHs with masses above $2\times10^8\,$g would be accessible to CMB-S4.

In the right panel of Fig.~\ref{fig:1_a_b}, we show the results for a benchmark extended distribution from an EMDE with $\left\langle a^* \right\rangle \simeq 0.64$, along with
a monochromatic distribution with $a^* = 0.64$ (more details on this are discussed in Section~\ref{emdeetc} and Appendix~\ref{sec:spin_dist_MD}).\footnote{The value $a^* = 0.63$ mentioned in~\cite{Harada:2017fjm} is the peak value of the distribution, not its average.} 
The spin distribution in the right panel of Fig.~\ref{fig:1_a_b} due to early matter domination is significantly different from that in the left panel due to inspirals. In particular, this EMDE spin distribution is less symmetric and much more broad than the inspiral distribution. The relative discrepancy between the extended distribution and the monochromatic distribution is therefore even greater ($\sim 60\%$) in the right panel than in the left panel of Fig.~\ref{fig:1_a_b}.  For this EMDE extended spin distribution, one finds that PBHs with masses above $\sim 7\times 10^7\,$g will, in fact, be probed by CMB-S4.  This conclusion stands in stark contrast to that inferred under the assumption of a monochromatic spin distribution at the peak or average spin.

The results for the extended spin distributions in both panels of Fig.~\ref{fig:1_a_b} are also shown  for the two prescriptions for calculating the reheating temperature, as discussed in  Section~\ref{sec:BlackHawk}; (1) instantaneous reheating at the evaporation time of the last PBH (with the lowest spin) is shown in grey, and (2) the weighted average PBH evaporation time using Eq.~\eqref{eq:average_tau} is shown in black.   In both panels, one can see that prescription (2) results in a shift in the $\DNeff$ curve to higher PBH mass relative to the results assuming prescription (1).  This can be understood on the basis of the reheating temperature from prescription (1) being smaller than the reheating temperature from prescription (2). Indeed, higher spin PBHs evaporate faster, and are better accounted for in prescription (2).  Thus, one could achieve the same reheating temperature (and therefore the same $\DNeff$) with prescription (1) by assuming a higher PBH mass.

\begin{figure}
    \centering
    \includegraphics[scale = 1]{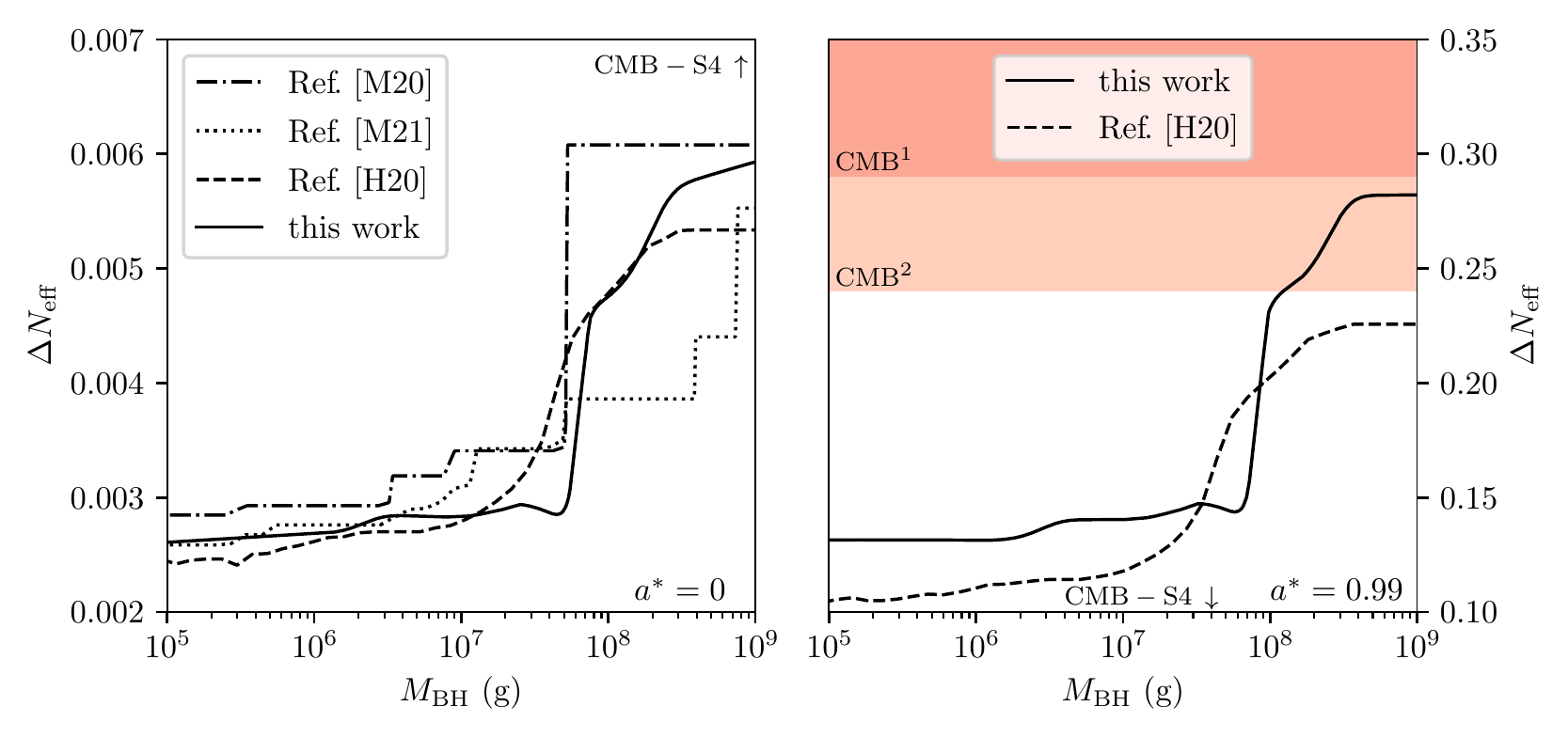}
    \caption{\textbf{Left:} Comparison between our precision calculation for $a^* = 0$ (solid) and the results of Ref.~[M20]~\cite{Masina:2020xhk} (dot-dashed) updated in Ref.~[M21]~\cite{Masina:2021zpu} (dotted) and Ref.~[H20]~\cite{Hooper:2020evu} (dashed).
    \textbf{Right:} Comparison between the precision calculation for $a^* = 0.99$ (solid) and the results of Ref.~[H20]~\cite{Hooper:2020evu} (dashed).
    The 95\% C.L. CMB limits on $\Delta N_\text{eff}$ (shaded areas) are taken from~\cite{Aghanim:2018eyx} (CMB$^1$: TT+low E, CMB$^2$: TT,TE,EE+low E).}
    \label{fig:1_c}
\end{figure}

In Fig.~\ref{fig:1_c}, we compare the values of $\DNeff$ obtained with precision calculations using \texttt{BlackHawk} to recent calculations in the literature. In the left panel of Fig.~\ref{fig:1_c}, we consider $a^*=0$ (Schwarzschild), and compare the $\DNeff$ from \texttt{BlackHawk} (solid) with those calculated in Refs.~\cite{Hooper:2020evu} (denoted as [H20], dashed) and Ref.~\cite{Masina:2020xhk} (denoted as [M20], dot-dashed) updated in Ref.~\cite{Masina:2021zpu} (denoted as [M21], dotted) with the use of \texttt{BlackHawk}. The relative discrepancies in these cases are  $\sim 10\%$.  In the right panel of Fig.~\ref{fig:1_c}, we consider $a^*=0.99$, and compare with the results of Ref.~\cite{Hooper:2020evu}, where we find a $\sim 20\%$ discrepancy.
As discussed in Section~\ref{sec:reheat}, an important difference between our results (solid) and other calculations in the literature is that here we take the values for $g_{*}(T)$ and $g_{*,S}(T)$ tabulated in the public code \texttt{SuperIso Relic}~\cite{Arbey:2009gu,Arbey:2018msw}.
Near $M\mrm{BH} \sim 8\times 10^{7}\,$g (corresponding to $T\mrm{RH} \sim 100\,$MeV), the number of degrees of freedom is very sensitive to the QCD equation of state, and the precise behavior of $\DNeff$ is evident.
That said, using a simple step function for $g_{*}(T)$ and $g_{*,S}(T)$ gives results qualitatively similar to ours~\cite{Masina:2020xhk,Masina:2021zpu}. 
This precision calculation reveals that highly spinning PBHs with $a^* = 0.99$ and masses $M\mrm{BH}\gtrsim 2\times 10^{8}\,$g that dominated the Universe before BBN are, in fact, {\it already excluded} by CMB$^2$ constraints on $\DNeff$~\cite{Aghanim:2018eyx}.

\subsection{Early matter domination and extremal spins}\label{emdeetc}

In this subsection, we present the results of precision calculation of $\DNeff$ for PBH spin distributions from a period of early matter domination. This is the first time a prediction for $\DNeff$ from PBHs produced during an EMDE has been calculated.  Here, we assume that PBHs produced during the EMDE come to dominate the Universe by the time of Hawking evaporation. 
The validity of this assumption depends on the physics behind EMDE; as an example,  we consider the conditions under which this happens when an EMDE is caused by a gravitationally coupled modulus field in Appendix~\ref{emdemodel}.

We use the spin distributions from Ref.~\cite{Harada:2017fjm} as benchmarks, the details of which are discussed in Appendix~\ref{sec:spin_dist_MD}. Angular momentum within a comoving region of space has two components; the first-order contribution (``the first-order effect") originating from deviation of the boundary of the volume from a sphere, and the second-order contribution (``the second-order effect") sourced by density fluctuations in the comoving region (for a detailed treatment, we refer to ~\cite{Harada:2017fjm}). 
The first-order effect usually dominates (when the initial deviation of the boundary of collapsing region from a sphere is large), but an almost spherical initial collapsing region can diminish the first-order effect and make the second-order effect the dominant one.

\begin{figure}
    \centering
    \includegraphics[scale = 1]{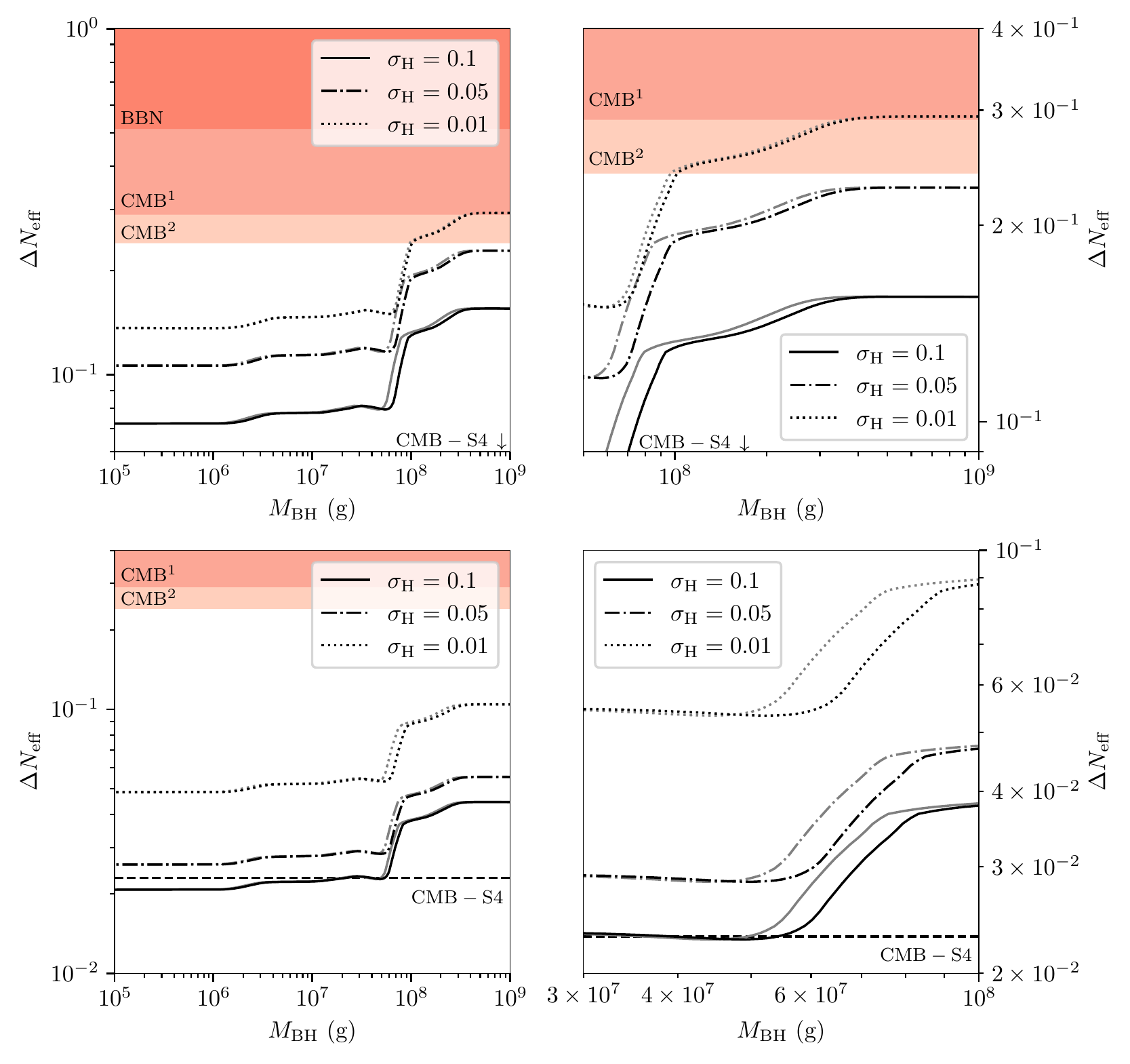}
    \caption{\textbf{Upper left:} $\DNeff$ results for PBH distributions formed during an early matter domination era due to first order effect~\cite{Harada:2017fjm} with $\sigma_\text{H} = \lbrace 0.01, 0.05, 1\rbrace$ (dotted, dot-dashed and solid respectively). \textbf{Upper right:} A zoom into the region where $\DNeff$ is constrained by current CMB limits.
    \textbf{Lower left:} $\DNeff$ results for PBH distributions formed during an early matter domination era due to second order effect~\cite{Harada:2017fjm} with $\sigma_\text{H} = \lbrace 0.01, 0.05, 1\rbrace$ (dotted, dot-dashed and solid respectively). \textbf{Lower right:} A zoom into the region where $\DNeff$ shows brutal change due to the step shape of $g_*(T\mrm{RH})$.
    The black (grey) curves correspond to instantaneous reheating at the weighted average value of the black hole lifetimes (last black hole evaporation). The 95\% C.L. limits on $\Delta N_\text{eff}$ from CMB (shaded areas) are taken from~\cite{Aghanim:2018eyx} (CMB$^1$: TT+low E, CMB$^2$: TT,TE,EE+low E); the 95\% limit from BBN is computed in Section~\ref{sec:BBN}; the prospective CMB-S4 constraint (horizontal dashed line) is extracted from~\cite{Hooper:2020evu}.}
    \label{fig:2_b}
\end{figure}

In Fig.~\ref{fig:2_b}, we present $\DNeff$ results for PBHs formed during an EMDE, with spin distributions due to first- and second-order effects in the upper and lower panels, respectively.   In each panel, we show results for three different values of $\sigma_{\rm{H}}$, the mean variance of the density perturbations at horizon entry
\begin{equation}
    \sigma_{\rm{H}} = \langle \delta_{s} (t_{\rm{H}})^2\rangle^{1/2}\,,
\end{equation}
where $\langle\delta_s^2\rangle$ is the variance of the density perturbations integrated over the volume of a sphere and $t_{\rm H}$ is the time of horizon entry (for further details, see Appendix~\ref{sec:spin_dist_MD}). $\sigma_{\rm H}$ controls the shape of the spin distribution, as well as the peak location.  For both the first- and second-order effects, larger 
 $\sigma_{\rm H}$ leads to more broad spin distributions.  Increasing $\sigma_{\rm H}$ also shifts the peak of the second-order distribution away from $a^*=1$ to smaller values of $a^*$.  As mentioned in Section~\ref{sec:comparison}, for $\sigma_{\rm H}=0.1$, the peak average of the spin distribution from the second-order effect is located at $a^*=0.64$. Note that either the first- or second-order effects could dominate, as discussed in Appendix~\ref{sec:spin_dist_MD}.

The upper panels of Fig.~\ref{fig:2_b} show $\DNeff$ due to spin distributions dominated by the first-order effect.  We see that for $\sigma_{\rm H}$ small enough, the largest PBH masses are \textit{already excluded} by CMB$^2$, and in some cases even CMB$^1$. In the upper right panel, we see in detail that for $\sigma_{\rm H} \lesssim 0.01$, $M_{\rm BH} > 10^8\,$g are excluded by CMB$^2$ constraints. For EMDE spin distributions dominated by the first order effect, the entire PBH mass range $10^{-1} \,{\rm g} < M_{\rm BH} < 10^9\,$g will be probed by CMB Stage 4.

It is clear from the lower panels of Fig.~\ref{fig:2_b} that PBH spin distributions from an EMDE are not constrained by current CMB or BBN limits on $\DNeff$ if the spin distribution is dominated by the second-order effect. However, these would be probed by CMB Stage 4 measurements. We can see in the lower left panel of Fig.~\ref{fig:2_b} that for $\sigma\mrm{H}$ small enough ($\lesssim 0.1$), all PBH masses in the range $10^{-1}-10^{9}\,$g will be probed by CMB Stage 4. For the value $\sigma\mrm{H} = 0.1$, only PBHs in the high mass end of this range $3\times10^{7}-10^{9}\,$g will be accessible to CMB Stage 4.

Another noticeable feature in all panels of Fig.~\ref{fig:2_b} is the shift of $\DNeff$ towards higher PBH masses if one takes reheating time as the average weighted lifetime $\langle \tau \rangle$ (black curves) compared to the time of evaporation of the last PBH (grey curves). This is consistent with what was observed in  Fig.~\ref{fig:1_a_b}. This shift is most sizeable for extremal spin distribution for which the average spin is $\langle a^* \rangle \sim 1$, \textit{i.e.}~small $\sigma_{\rm H}$. This is especially clear in the lower right panel of Fig.~\ref{fig:2_b}, which zooms in to the region of strong variation of $\DNeff$ in the lower left panel. The difference in these results due to the different prescriptions for reheating time particularly
affects $\DNeff$ in the mass range $M\sim 5-9\times10^{7}\,$g. This is also the region where $\DNeff$ is most affected by the precise shape of $g_{*}(T)$ and $g_{*,S}(T)$.

\begin{figure}
    \centering
    \includegraphics[scale = 1]{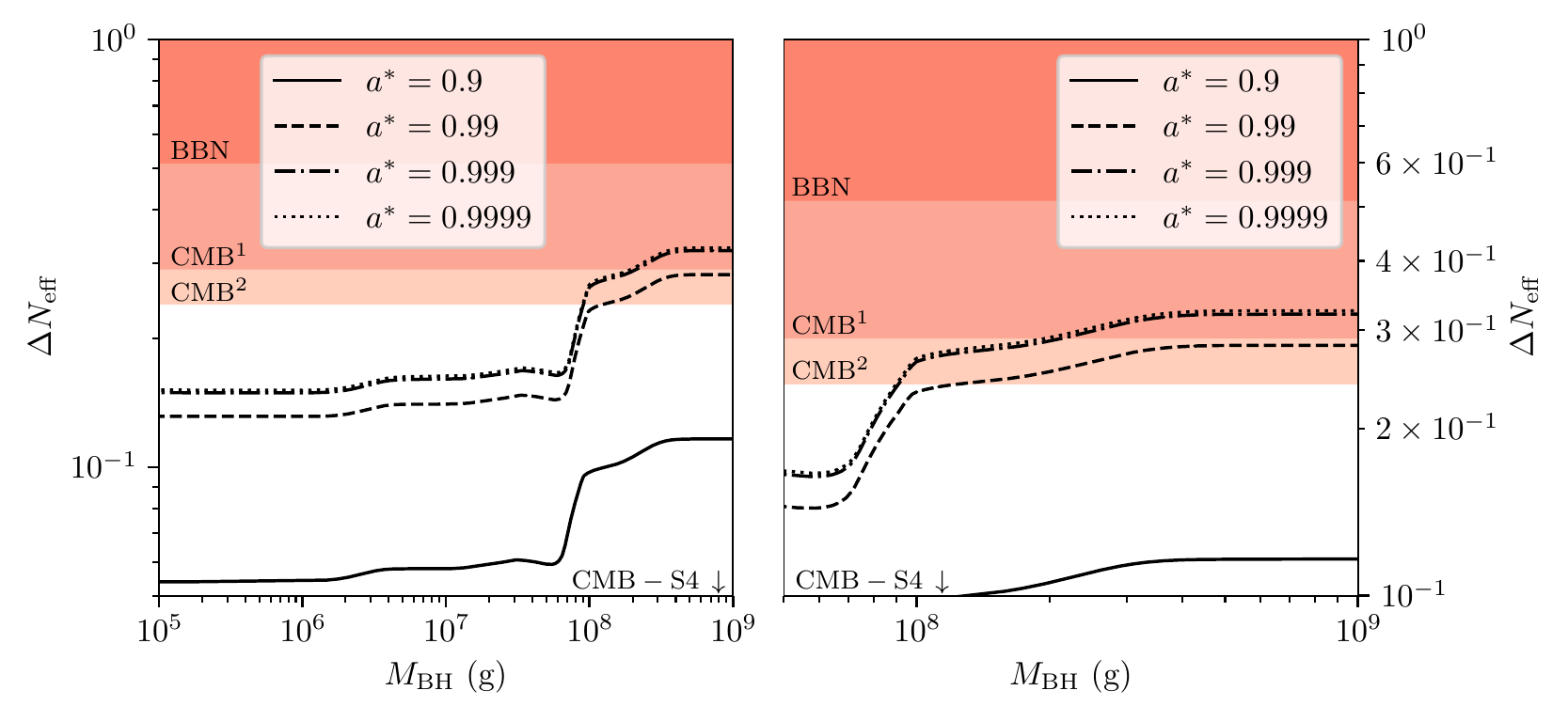}
    \caption{\textbf{Left:} Our results for high spin PBHs with $a^* = \lbrace 0.9, 0.99, 0.999, 0.9999 \rbrace$ (solid, dashed, dot-dashed and dotted respectively). The last two curves are difficult to distinguish. \textbf{Right:} A zoom-in of the CMB exclusion region for highly spinning massive PBHs. The 95\% C.L. limits on $\Delta N_\text{eff}$ from CMB (shaded areas) are taken from~\cite{Aghanim:2018eyx} (CMB$^1$: TT+low E, CMB$^2$: TT,TE,EE+low E); the 95\% limit from BBN is computed in Section~\ref{sec:BBN}; the prospective CMB-S4 constraint (horizontal dashed line) is extracted from~\cite{Hooper:2020evu}.}
    \label{fig:2_c}
\end{figure}

We next turn to an investigation of $\DNeff$ for near-extremal PBH spins. 
In Fig.~\ref{fig:2_c}, we present $\DNeff$ for monochromatic spin distributions with $a^*\gtrsim 0.9$, under the assumption that the PBHs dominate the energy density of the Universe before BBN. As in Fig.~\ref{fig:1_c}, we see that these highly spinning PBHs are \textit{already excluded} by current CMB$^2$ constraints for large enough PBH masses. We also see that the excluded mass range grows as the spin increases, due to the 
 shorter lifetime of spinning PBHs. Furthermore, for the largest PBH spins we consider, the increase in $\DNeff$, which is due to the enhanced emission of high spin particles (spin 2 most of all), saturates. Indeed, the Hawking emissivity of near extremal PBHs does not grow to infinity as $a^*\rightarrow 1$ but instead saturates. 

To be specific, we see in Fig.~\ref{fig:1_c}  that 
the future CMB Stage 4 measurements will be sensitive to extremal values of PBH spins $a^* \gtrsim 0.9$.
From the right panel of Fig.~\ref{fig:2_c} it is evident that PBHs with spin $a^* \gtrsim 0.99$ and mass $M\mrm{BH} \gtrsim 10^8\,$g are excluded by the CMB$^2$ stringent constraints. PBHs with even higher spin $a^* \gtrsim 0.999$ are constrained by the CMB$^1$ conservative constraints, but only for masses $M\mrm{BH} \gtrsim 2\times 10^8\,$g. This is, to our knowledge, the first constraints put on light spinning PBHs from $\DNeff$ from \textit{current} CMB limits.

\begin{figure}
    \centering
    \includegraphics[scale = 1]{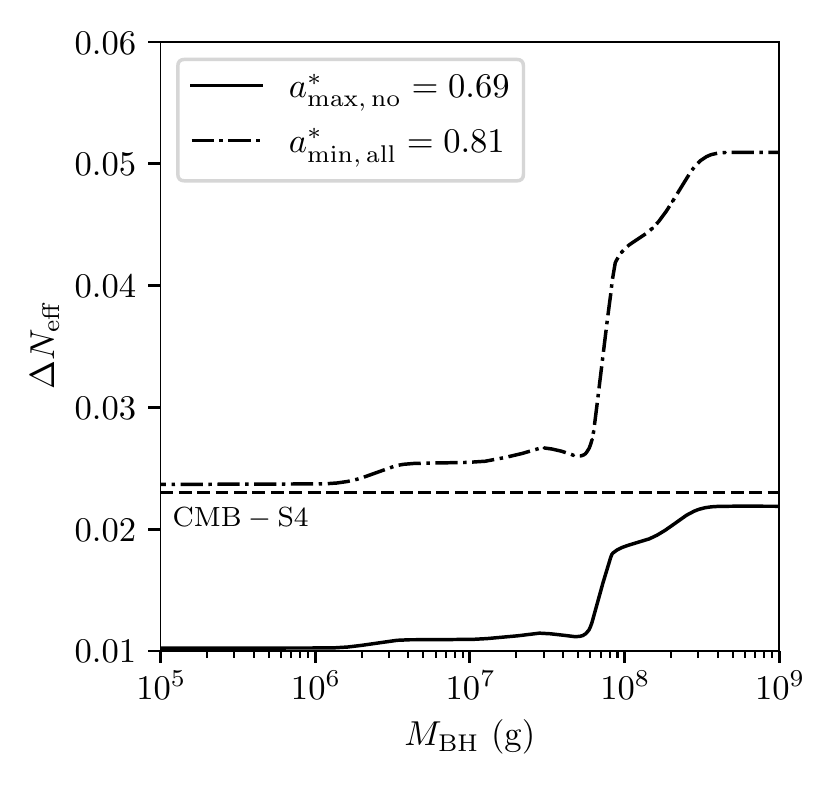}
    \caption{Low and high cut-off values for CMB Stage 4 exclusion, corresponding to $a\mrm{max,\,no}^* = 0.69$ and $a\mrm{min,\,all}^* = 0.81$ (solid and dot-dashed respectively). The prospective CMB-S4 constraint (horizontal dashed line) is extracted from~\cite{Hooper:2020evu}.}
    \label{fig:2_d}
\end{figure}

Finally, we explore the capability of the CMB Stage 4 experiment to explore PBHs with monochromatic spins, under the assumption that PBHs dominated the energy density of the Universe prior to BBN.  In Fig.~\ref{fig:2_d}, we present the smallest monochromatic spin for which CMB Stage 4 will be sensitive to the {\it entire} mass range considered here, $a^*\mrm{min,\,all}$, as well as the largest monochromatic spin for which CMB Stage 4 will not be sensitive to any part of the mass range, $a^*_{\rm max,\,no}$. 
We find that the smallest monochromatic spin for which CMB Stage 4 will be sensitive to the whole range of masses is $a^*\mrm{min,\,all} \simeq 0.81\,$. For a monochromatic spin distribution with $a^* > a^*\mrm{min,\,all}$, CMB Stage 4 will probe all PBH masses $10^{-1}\,{\rm g} < M_{\rm BH}<10^9\,$g.  On the other hand, the smallest monochromatic spin value for which CMB Stage 4 can constrain any of the PBH masses is
$a^*\mrm{max,\,no} \simeq 0.69$.  For $a^*\lesssim a^*\mrm{max,\,no}$, the entire mass range would be inaccessible to CMB Stage 4, while for $a^*\gtrsim a^*\mrm{max,\,no}$ only the heaviest PBHs ($M\mrm{BH}\sim 10^{9}\,$g) will be probed.

While the results in Fig.~\ref{fig:2_d} apply to monochromatic spin distributions, the same question can in principle be answered for various types of extended spin distributions.  As discussed in Section~\ref{sec:comparison} and demonstrated in Fig.~\ref{fig:1_a_b}, one can expect a $\sim 25-60\%$ relative discrepancy between the $\DNeff$ prediction for monochromatic spin distributions relative to the extended distributions we consider here. Indeed, for a scenario such as early matter domination, which induces a particular spin distribution for PBHs, one could even explore the range of cosmological parameters that yield $\DNeff$ to which next generation experiments will be sensitive.

\section{Effect on BBN}
\label{sec:BBN}

In this Section, we outline how the dark radiation yield from light PBH evaporation can affect BBN. If PBHs evaporate before the onset of BBN ($t_\text{eva} \lesssim 1\,$s or $M_{\rm BH}\lesssim 10^{9}\,$g), then the emitted SM particles thermalize to the expected plasma density and provide no measurable effect on BBN. The dark sector, which is also emitted by Hawking radiation, however, provides an additional source of density in the Friedmann equations compared to standard BBN. This sector does not interact with the SM, thus its temperature is decoupled from the plasma temperature. However, the dark radiation can be treated as an additional effective number of neutrinos $N\mrm{eff}$ during BBN and up to the time of photon decoupling. Thus, the $\Delta N_\text{eff}$ constraints from BBN can be used to constrain the dark radiation density before BBN. 

We use the public code \texttt{AlterBBN} \cite{Arbey:2011nf,Arbey:2018zfh}, which computes the abundances of the light chemical elements in alternative cosmological scenarios, such as with the addition of a dark radiation density. Comparison with the fiducial values for these abundances, in particular $^2$H and $^4$He measured in old gas clouds, provides constraints on $\Delta N_\text{eff}$.

\begin{figure}
    \centering
    \includegraphics[scale = 1]{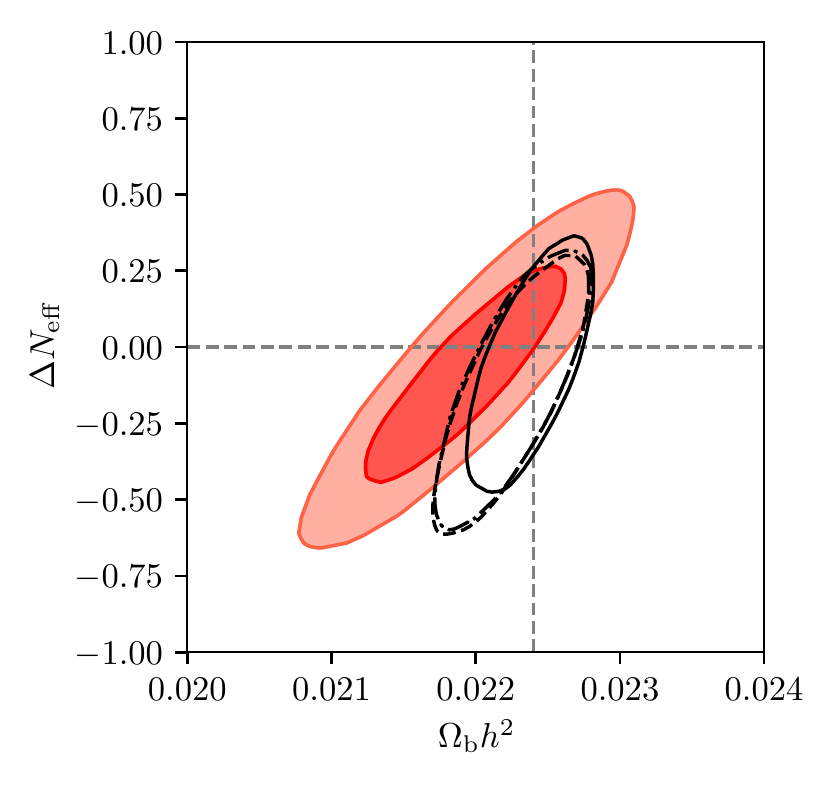}
    \caption{Color contours: 68\% and 95\% C.L. regions obtained using the $^2$H and $^4$He BBN constraints, as recomputed with \texttt{AlterBBN} for this work. Lines: 95\% C.L. obtained by the Planck Collaboration~\cite{Aghanim:2018eyx} (dot-dashed: TT,TE,EE+lowE, dashed: TT,TE,EE+lowE+lensing, solid: TT,TE,EE+lowE+lensing+BAO), with the vertical and horizontal dashed grey lines marking the standard values of the parameters $\Omega\mrm{b} h^2 = 0.0224$ and $\DNeff = 0$, respectively.}
    \label{fig:Neff}
\end{figure}

The master parameter for BBN is the baryon-to-photon ratio $\eta$, which is related to the reduced baryon cosmological parameter $\Omega\mrm{b} h^2$ via
\begin{equation}
    \eta = \frac{n\mrm{b}}{n_\gamma} = \frac{3\pi M\mrm{Pl}^2 k^2}{m_{\rm b} 16\zeta_3 T_0^3} \Omega\mrm{b} h^2 = 274\times 10^{-10} \;\Omega\mrm{b} h^2 \,,
\end{equation}
where $k=100\,\text{km/s/Mpc}$ is the Hubble parameter scale today, $M\mrm{Pl}$ the Planck mass, $m\mrm{b}$ is the average baryon mass, and $T_0=2.7255\,$K is the CMB temperature today. The constraints on $\DNeff$ are computed with $\Omega\mrm{b}h^2$ as a free parameter. Its central value is $\Omega\mrm{b}h^2 = 0.0224$~\cite{Aghanim:2018eyx}. Inside \texttt{AlterBBN}, the observational values of the chemical element abundances used to obtain the updated $\DNeff$ constraints are
\begin{eqnarray}
Y\mrm{P} &=& 0.2453 \pm 0.0034,\;\; \text{\cite{Aver:2020fon}},\\
{\rm D/H} &=& (2.527 \pm 0.030)\times 10^{-5}, \;\;\text{\cite{Cooke:2017cwo}}.
\end{eqnarray}
The improved nuclear rate for ${\rm D} + {\rm p} \to\, ^3{\rm He} + \gamma$ by LUNA has been implemented into the code~\cite{Mossa:2020gjc}.

In Fig.~\ref{fig:Neff}, we present the BBN constraints on $\DNeff$ with $\Omega\mrm{b}h^2$ as a free parameter. The light and dark shaded red regions correspond to the 68\% and 95\% confidence level regions obtained using the $^2$H and $^4$He BBN constraints, as recomputed with \texttt{AlterBBN} for this work (these are the BBN constraints used in Figs.~\ref{fig:2_b} and~\ref{fig:2_c}).  The dot-dashed, dashed, and solid contours correspond to the 95\% confidence level regions obtained by the Planck Collaboration~\cite{Aghanim:2018eyx} (dot-dashed: TT,TE,EE+lowE, dashed: TT,TE,EE+lowE+lensing, solid: TT,TE,EE+lowE+lensing+BAO), with the vertical and horizontal dashed grey lines marking the standard values of the parameters $\Omega\mrm{b} h^2 = 0.0224$ and $\DNeff = 0$, respectively. 

If PBHs evaporate during or after BBN, then the effects are much more complicated and require  careful treatment, beyond the scope of the current study. First, the energetic hadronic emission just before BBN can trigger $\text{p}\,\longleftrightarrow\,\text{n}$ interconversion and thus modify the $\text{p}/\text{n}$ ratio at the beginning of BBN. This ratio strongly affects the final $^4$He abundance and is thus severely constrained. Second, hadronic injection (mesons) during BBN can trigger nuclear reactions through hadrodissociation and can modify the abundance of intermediate light elements. This may modify the final $^2$H abundance and is also severely constrained. Third and last, the emission of energetic photons at the end of BBN can still destroy BBN products through photodissociation and thus can modify the final abundances before recombination. All these phenomena are associated with the evaporation of $M\gtrsim 10^{9}\,$g PBHs. We refer the interested reader to \cite{Zeldovitch1977,Sedelnikov1996,Liddle:1998nt,Kohri:1999ex,Carr:2009jm,Carr:2020gox,Keith:2020jww,Luo:2020dlg} for detailed analyses of these.

\section{CONCLUSIONS}
\label{sec:conclusion}

Our purpose in this paper has been to conduct precision studies of dark radiation emanating from spinning PBHs. We have concentrated on the case of gravitons. Our precision study incorporated spin distributions of PBHs and a careful treatment of the reheating temperature and relativistic degrees of freedom.  We studied the impacts of each of these three precision elements on the calculation of $\DNeff$ due to graviton emission from PBHs, and applied the calculation to a scenario with extended PBH spin distributions due to an early matter dominated era (EMDE).

There are two main effects related to incorporating extended PBH spin distributions relative to monochromatic spin distributions.  First, since a BH's lifetime is related to its spin, a spin distribution will result in a distribution of evaporation times.  The second, dominant, effect is that PBHs with high spins emit more particles with higher spins, i.e.~gravitons.  So a spin distribution that extends to higher spins will result in more graviton emission
relative to a corresponding monochromatic spin approximation, and thus a larger prediction for $\DNeff$.

In undertaking a precision study, we find that it is also important to consider a precise formulation for the number of relativistic degrees of freedom as a function of temperature. We show that different characterizations for the degrees of freedom lead to different conclusions regarding experimental sensitivity to various models.  In fact, for PBHs with masses $M_{\rm BH}\gtrsim {\rm few}\times10^7\,$g that dominated the Universe before BBN, one finds very different predictions for $\DNeff$.
Different prescriptions for the reheating temperature due to PBH evaporation also lead to variations of $\DNeff$.  These are relatively small in comparison to the other effects considered, but careful attention to the reheating temperature is relevant to make a precise statement regarding experimental sensitivity for some PBH masses.

Our main application was to study gravitons coming from Hawking evaporation of PBHs created during an EMDE. If such PBHs come to dominate the Universe prior to final evaporation, the resulting dark radiation can be probed by current BBN and CMB constraints, as well as future CMB Stage 4 experiments. We have found that   PBHs with spin distribution due to the first-order effect are constrained by current CMB bounds on $\DNeff$ in the mass range $10^{8}-10^{9}\,$g, and would be completely constrained in the mass range $10^{-1}-10^{9}\,$g by CMB Stage 4 projections. 
PBHs formed during an EMDE with spin distribution due to the second-order effect, while not constrained by current BBN or CMB bounds on $\DNeff$, would be completely constrained in the mass range $10^{-1}-10^{9}\,$g by CMB Stage 4 experiments for all scenarios except for the largest $\sigma_{\rm H}$ considered here. In terms of the modulus sector, we found that for a variety of PBH spin distributions and fractions $\beta$ that have been considered in the literature, moduli with masses larger than $\sim 10^8$ GeV will be constrained by CMB Stage 4 experiments.

We also explored $\DNeff$ for near-extremal PBH spins. We find that if PBHs with monochromatic spin distributions with $a^*\gtrsim 0.99$ dominate the energy density of the Universe before BBN, current CMB constraints exclude PBHs with masses $m_{\rm BH} \gtrsim 10^8\,$g. As the spin increases toward 1, $\DNeff$ increases until it saturates, since the Hawking emissivity of near extremal PBHs does not grow to infinity as spin approaches 1 but instead saturates.  We therefore find that for increasing $a^*$ the minimal PBH mass excluded by current CMB measurements is shifted to lower PBH masses until saturation.   We also find that for PBHs with monochromatic spins $a^* >0.81$ that dominated the energy density of the Universe prior to BBN, all PBH masses in the range $10^{-1} \, {\rm g} < M_{\rm BH} <10^9\,$g will be probed by CMB Stage 4 experiments.

\textbf{Note:} Near the completion of this work, the authors became aware of the publication of Ref.~\cite{Masina:2021zpu}, where the author considers  Hawking radiation of light Kerr PBHs in the early Universe, in the mass range $10^{-5}-10^9\,$g. 
Ref.~\cite{Masina:2021zpu} considers the emission of light dark matter particles by Kerr PBHs, as an extension of the results of \cite{Auffinger:2020afu} for Schwarzschild PBHs, as well as the effect of emission of dark radiation by light Kerr PBHs, as considered in this work. We compare the results in \cite{Masina:2021zpu} to ours and others in the literature in subsection~\ref{sec:comparison}.


\appendix

\section{Early matter dominated eras and PBH  spin distributions} 

In this Appendix, we discuss possible PBH spin distributions $n(a^*)$ that are motivated by early Universe cosmology. These distributions will then be used in Eq.~\eqref{eq:ratio} to obtain $f_{\rm DR}$. We will focus mainly on two benchmark scenarios: a period of early matter domination, possibly by a string modulus and scenarios in which PBHs acquire spin by inspirals.

\subsection{PBH formation during an early matter dominated era and subsequent evaporation} \label{emdemodel}

In usual studies of an early matter domination era (EMDE) phase, the scenario is the following: after inflationary reheating, the Universe is filled with radiation and a modulus field, $\phi$. We will be agnostic about the origins of $\phi$ -- it could be a string modulus. We will assume that it couples to other fields via gravity only. Under fairly general assumptions, it is possible that $\phi$ is displaced from the minimum of its potential and starts to oscillate. Since energy density of modulus field redshifts 
like energy density of matter, it eventually dominates the energy density of the Universe and causes a transition from a radiation-dominated era to a matter-dominated era. During this modulus-dominated epoch, spinning PBHs can form. Modulus field will finally decay into radiation, reheat the Universe for a second time and give rise to a radiation-dominated era. Since energy density of PBHs also redshifts like matter, they can eventually dominate over radiation and lead to a matter-dominated epoch. In this case, after evaporation, their contribution to $\Delta N_{\rm eff}$ is not negligible. 

\begin{figure}
    \centering
    \includegraphics[scale = 1]{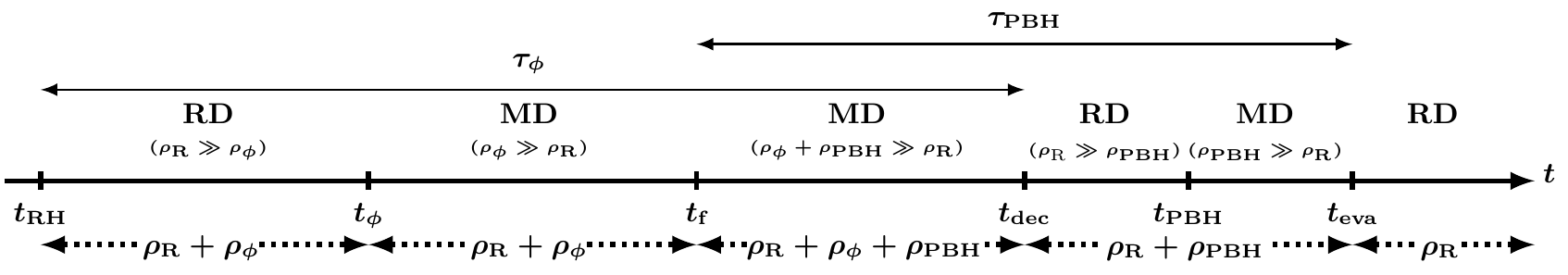}
    \caption{Thermal history of Universe reheated at $t=t_\text{RH}$, includes radiation (R), and a modulus field ($\phi$). After reheating radiation dominates energy density of the Universe, so Universe experiences a radiation-dominated (RD) epoch until energy density of $\phi$ dominates at $t=t_\phi$ and initiates a matter-dominated (MD) era. During this modulus-dominated era, spinning PBHs may form at $t=t_\text{f}$. At $t=t_\text{dec}$, $\phi$ stops to oscillate around its minimum and decays into radiation. After $t=t_\text{dec}$ Universe enters a RD epoch which may lead to a MD era at $t_\text{PBH}$ at which energy density of PBHs takes over if the lifetime of PBHs is long enough. PBHs will eventually deposit their energy content into the thermal bath at $t=t_\text{eva}$ due to Hawking evaporation. This is the onset of another RD epoch which will continue until radiation-matter equality time. Time intervals in the plot are for demonstrative purposes only and not indications of actual times.}
    \label{fig:timeline}
\end{figure}

To evaluate the initial abundance of PBHs for which a once modulus-dominated Universe may lead to a PBH-dominated epoch, one needs to trace back the evolution of energy density of each component to the onset of modulus-dominated era (see Fig.~\ref{fig:timeline}). We assume that following the reheating of Universe at $t=t_\text{RH}$, energy density of the modulus field becomes comparable with energy density of radiation at $t=t_\phi$, \textit{i.e.}, $\rho_\text{R}(t_\phi)\simeq\rho_\phi(t_\phi)$. Afterward a fraction $\beta$ of the total energy density of the Universe collapses into PBHs at $t=t_\text{f}$, \textit{i.e.}, $\beta\equiv\rho_\text{PBH}(t_\text{f})/\rho_\text{tot}(t_\text{f})$. Subsequently, the modulus field decays instantaneously into radiation at $t=t_\text{dec}$, and eventually PBHs evaporate at time $t=t_\text{eva}$.  
Then, to guarantee a PBH-dominated era, we need to make sure that at some time $t=t_\text{PBH}$, where $t_\text{dec}\lesssim t_\text{PBH}\lesssim t_\text{eva}$, we have $\rho_\text{PBH}(t_\text{PBH})\simeq \rho_\text{R}(t_\text{PBH})$. This leads to 
\begin{equation}
    \frac{a(t_\text{PBH})}{a(t_\text{dec})}=\frac{1-\beta}{\beta}\left[\frac{\frac{a(t_\phi)}{a(t_\text{dec})}+1}{\frac{a(t_\phi)}{a(t_\text{f})}+1}\right]\simeq \frac{1}{\beta}\,,
\end{equation}
where $a$ is the scale factor, and we assume that $\beta\ll1$. Since $t_\phi< t_\text{f}<t_\text{dec}$, ignoring $a(t_\phi)/a(t_\text{dec})$ and $a(t_\phi)/a(t_\text{f})$ can cause an overestimation up to a factor of 2. Demanding $t_\text{PBH}\lesssim t_\text{eva}$ (or equivalently $a(t_\text{PBH})\lesssim a(t_\text{eva})$) provides a lower bound on $\beta$ given by
\begin{equation}
    \beta \gtrsim \beta_c\equiv \frac{a(t_\text{dec})}{a(t_\text{eva})}=\frac{a(t_\text{RH}+\tau_\phi)}{a(t_\text{f}+\tau_\text{PBH})}=\sqrt{\frac{t_\text{RH}+\tau_\phi}{t_\text{f}+\tau_\text{PBH}}}\simeq \sqrt{\frac{\tau_\phi}{\tau_\text{PBH}}}\simeq\sqrt{\frac{M_\text{pl}^2}{m_\phi^3\tau_\text{PBH}}}\,,
    \label{eq:betac}
\end{equation}
where $\tau_\phi$ and $\tau_\text{PBH}$ are the lifetimes of modulus field and PBHs respectively. To evaluate Eq.~\eqref{eq:betac}, we use the fact that since $\beta_c$ corresponds to the case that PBHs dominated energy density almost at the time of their evaporation, between decay of the modulus field and evaporation time, $t_\text{dec}\lesssim t \lesssim t_\text{eva}$, the Universe undergoes a radiation-dominated stage. We also use $t_\text{RH}\ll \tau_\phi$, $t_\text{f}\ll \tau_\text{PBH}$, and $\Gamma_\phi\simeq m_\phi^3/M_\text{Pl}^2$.

The initial abundance of PBHs, $\beta$, that formed during a matter-dominated epoch and gained angular momentum due to the first- and second-order effects (see Subsection~\ref{sec:spin_dist_MD} for details), is calculated as a function of the mean variance of density perturbations at horizon entry, $\sigma_\text{H}$, by Ref.~\cite{Harada:2017fjm}. A certain value of $\sigma_\text{H}$ can give rise to a PBH-dominated epoch if $\beta(\sigma_\text{H})\gtrsim \beta_c$, or equivalently if
\begin{equation}
    m_\phi\gtrsim\frac{1}{\beta(\sigma_\text{H})^{2/3}}\left(\frac{M_\text{Pl}^2}{\tau_\text{PBH}}\right)^{1/3}.
    \label{eq:mphic}
\end{equation}
For sufficiently small black holes ($M_{\rm BH} \lesssim 10^{10}\,$g), the lifetime of a spinning black hole can be expressed as~\cite{Arbey:2019jmj}
\begin{equation}
    \tau_\text{BH}=c(\langle a^*\rangle)\frac{M_\text{BH}^3}{M_\text{Pl}^4}
    \sim \mathcal{O}\left(10^{-28}\, \text{s}\right) \left(\frac{M_{\rm BH}}{1\,\text{g}}\right)^3,
    \label{eq:tauBH}
\end{equation}
where $c(\langle a^*\rangle)$ depends on the average of the spin of the black hole, so here is a function of $\sigma_\text{H}$, and is calculated by \texttt{BlackHawk}. 

By combining Eqs.~(\ref{eq:mphic}) and (\ref{eq:tauBH}), we obtain
\begin{equation}
    m_\phi\gtrsim\frac{1}{\beta(\sigma_\text{H})^{2/3}}\frac{1}{c(\sigma_\text{H})^{1/3}}\frac{M_\text{Pl}^2}{M_\text{PBH}}\,,
\end{equation}

where for $\beta(\sigma_\text{H})$ we follow the numerically calculated curves in Fig.~5 of Ref.~\cite{Harada:2017fjm}. The authors have checked that the following semi-analytic formulae reproduce the behavior:
\begin{equation}
 \beta_1(\sigma_\text{H}) \simeq  \left\{
        \begin{array}{ll}
            3.244\times 10^{-14} \dfrac{q^{18}}{\sigma_\text{H}^4} \text{exp}\left[-0.004608\dfrac{q^4}{\sigma_\text{H}^2}\right] & \quad \sigma_\text{H}\lesssim 0.04, \\
            0.05556 \sigma_\text{H}^5& \quad 0.04\lesssim \sigma_\text{H}\lesssim 0.2\,,
        \end{array}
    \right.
\end{equation}
\begin{equation}
 \beta_2(\sigma_\text{H}) \simeq  \left\{
        \begin{array}{ll}
            1.921\times 10^{-7} \mathcal{I}^6\sigma_\text{H}^2 \text{exp}\left[-0.1474\dfrac{\mathcal{I}^{4/3}}{\sigma_\text{H}^{2/3}}\right] & \quad \sigma_\text{H}\lesssim 0.005, \\
            0.05556 \sigma_\text{H}^5& \quad 0.005\lesssim \sigma_\text{H}\lesssim 0.2\,,
        \end{array}
    \right.
\end{equation}
where $\mathcal{I}=1$ and $q=\sqrt{2}$. For details, we refer to Ref.~\cite{Harada:2017fjm}.

\begin{figure}
    \centering
    \includegraphics[scale = 0.6]{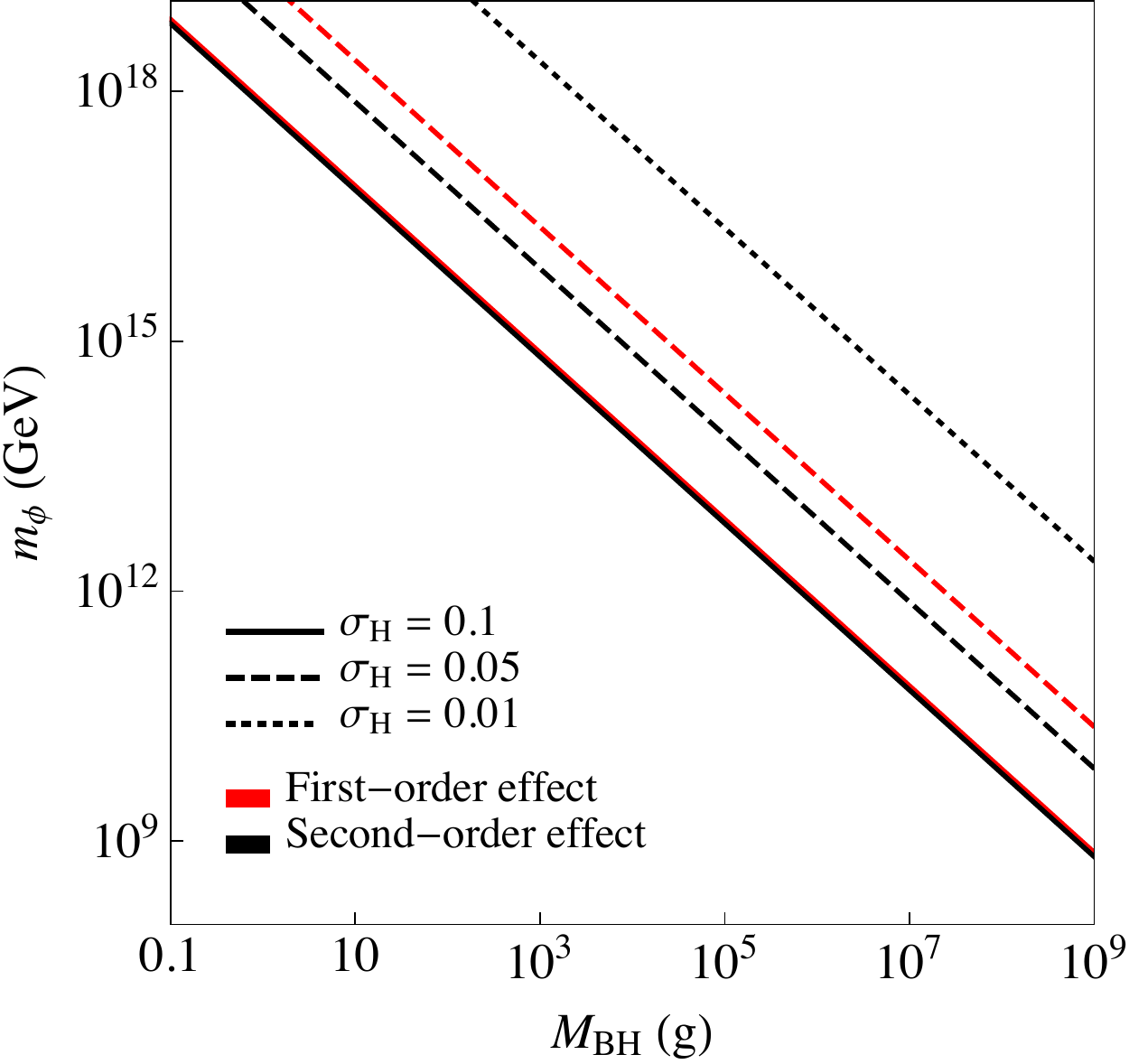}
    \caption{The lower bound on the mass of the modulus field which can later lead to a PBH-dominated era for benchmark values of $\sigma\mrm{H}$ that we use in this paper (dotted line for $\sigma\mrm{H}=0.01$, dashed line for $\sigma\mrm{H}=0.05$ and solid line for $\sigma\mrm{H}=0.1$), for both the first- (red) and second-order (black) effects. For the first-order effect, $\beta_c(0.01)$ is so small that the corresponding lower bound on the mass of the modulus field is always larger than $M_\text{Pl}$.} 
    \label{fig:order2fig}
\end{figure}

Fig.~\ref{fig:order2fig} displays the lower bound on the mass of the modulus field which can later lead to a PBH-dominated era for benchmark values of $\sigma\mrm{H}$ that we use in this paper, for both the first- and second-order effects.

\subsection{PBH spin distributions from an early matter dominated era}
\label{sec:spin_dist_MD}

In the early Universe, density fluctuations, $\delta=\delta\rho/\rho$, grow after they enter the cosmological horizon. In a radiation-dominated epoch, if density fluctuations are greater than  a threshold, they can collapse into a PBH with mass bounded by the total mass within the horizon. In a matter-dominated epoch, the absence or significant reduction of the pressure gradient force enhances PBHs formation rate and it is the deviation from spherical symmetry that governs the probability of PBH formation~\cite{Khlopov:1980mg}.

Since in cosmological perturbation theory, the rotational mode is not growing to linear order, the effect of rotation in the formation of PBHs is naively expected to be unimportant.  As a matter of fact, detailed calculation~\cite{Harada:2017fjm} shows that angular momentum plays a very important role in the formation of PBHs in the matter-dominated phase. Here we briefly review the importance of rotation in PBH formation in a matter-dominated epoch and the spin distribution of these PBHs by following the theory of angular momentum in structure formation adopted in~\cite{Harada:2017fjm}.

Angular momentum within a comoving region of space has two components; the first-order contribution originating from deviation of the boundary of the volume from a sphere which can be described by an ellipsoid, and  the second-order contribution sourced by density fluctuations in the comoving region. Assuming different modes carry random phases, the variance of the angular momentum within a sphere of comoving radius $r_0$ can be written as
\begin{equation}
\langle\textbf{L}^2\rangle=\langle\textbf{L}_{(1)}^2\rangle+\langle\textbf{L}_{(2)}^2\rangle\,,
\end{equation}
where 
\begin{equation}
\langle\textbf{L}_{(1)}^2\rangle^{1/2}\simeq\frac{2}{5\sqrt{15}}q\frac{MR^2}{t}\langle\delta_s^2\rangle^{1/2}\,,~~~~~\langle\textbf{L}_{(2)}^2\rangle^{1/2}\simeq\frac{2}{15}\mathcal{I}\frac{MR^2}{t}\langle\delta_s^2\rangle\,,
\end{equation}
and subscripts 1 and 2 represent the first- and second-order contributions respectively.

In the above expressions,  $M = (4\pi/3)\rho_0(a r_0)^3$ is the mass inside the spherical region of interest and $R=a r_0$ is the physical radius of the region, $q$ is the dimensionless parameter of the initial reduced quadrupole moment of the mass, $\mathcal{I}$ is of order unity, and  $\langle\delta_s^2\rangle\sim t^{4/3}$ is the variance of $\delta_s$, the density perturbation integrated over volume of the sphere. By normalizing them at the time of horizon entry, $t=t\mrm{H}$, we have
\begin{equation}
\langle\textbf{L}_{(1)}^2\rangle^{1/2}\simeq\frac{6}{5\sqrt{15}}q M^2\sigma\mrm{H}\left(\frac{t}{t\mrm{H}}\right),~~~~~\langle\textbf{L}_{(2)}^2\rangle^{1/2}\simeq\frac{2}{5}\mathcal{I}M^2\sigma\mrm{H}^2\left(\frac{t}{t\mrm{H}}\right)^{5/3},
\end{equation}
where $\sigma\mrm{H}\equiv \langle\delta_s(t\mrm{H})^2\rangle^{1/2}$.

The corresponding dimensionless angular momentum can be estimated as
\begin{equation}
\langle a_{(1)}^{*2}\rangle^{1/2}\simeq\frac{2}{5}\sqrt{\frac{3}{5}}q\sigma\mrm{H}\left(\frac{t}{t\mrm{H}}\right),~~~~~\langle a_{(2)}^{*2}\rangle^{1/2}\simeq\frac{2}{5}\mathcal{I}\sigma\mrm{H}^2\left(\frac{t}{t\mrm{H}}\right)^{5/3}.
\end{equation}
The value of angular momentum grows with time until nonlinearity becomes important. After this moment which is the time of maximum expansion, $t\mrm{max}$, linear perturbation theory is not valid any longer. The collapse of the overdense region begins and it becomes separated from the evolution of the Universe. Therefore after $t\mrm{max}$ angular momentum approaches a constant value. By demanding $\langle\delta_s(\langle t\mrm{max}\rangle)^2\rangle^{1/2}=1$, the average value of $t\mrm{max}$ can be estimated as $\langle t\mrm{max}\rangle \simeq t\mrm{H}\sigma\mrm{H}^{-3/2}$ and accordingly, the average value of the first- and second-order angular momenta are given by
\begin{equation}
\langle a_{(1)}^{*2}\rangle^{1/2}\simeq\frac{2}{5}\sqrt{\frac{3}{5}}q\sigma\mrm{H}^{-1/2}\,,~~~~~\langle a_{(2)}^{*2}\rangle^{1/2}\simeq\frac{2}{5}\mathcal{I}\sigma\mrm{H}^{-1/2}\,.
\label{eq:qeffect}
\end{equation}
The dominant component is chosen as the final angular momentum; $\langle a^{*2}\rangle^{1/2}\simeq \text{max} \left(\langle a_{(1)}^{*2}\rangle^{1/2},\langle a_{(2)}^{*2}\rangle^{1/2}\right)$. Only a minority of masses with $\langle a^{*2}\rangle^{1/2}\lesssim 1$ ($\sigma\mrm{H} \gtrsim 0.1$) can overcome centrifugal force and collapse directly to PBHs. Therefore, angular momentum strongly suppresses formation of PBHs and most of the PBHs are rapidly rotating at the time of formation. By comparing the first- and second-order angular momenta in Eq.~\eqref{eq:qeffect}, we see that the magnitude of $q$, which quantifies initial deviations of the collapsing region from a sphere, determines dominant effect; a large $q$ (large initial deviation from a sphere) leads to first-order dominance, on the other hand a small $q$ (an almost spherical initial collapsing region) makes the second-order effect the dominant one. 

In spite of the complicated dependence of angular momentum on different coupled modes, a hypothesis facilitates obtaining the distribution function for spins; since both $\langle \textbf{L}_{(1)}^2\rangle$ and $\langle \delta_s^2\rangle$ include self-coupling of single modes while $\langle \textbf{L}_{(2)}^2\rangle$ consists of the coupling of two independent modes which are not parallel to each other, it is reasonable to assume that $|\textbf{L}_{(1)}| \propto \delta_s$ and $|\textbf{L}_{(2)}| \propto \langle \delta_s^2\rangle^{1/2} \delta_s$ , or more precisely 
\begin{equation}
|\textbf{L}_{(1)}| \simeq\frac{2}{5\sqrt{15}}q\frac{MR^2}{t}\delta_s\,,~~~~~|\textbf{L}_{(2)}|\simeq\frac{2}{15}\mathcal{I}\frac{MR^2}{t}\langle\delta_s^2\rangle^{1/2}\delta_s\,.
\end{equation}
By using $t\mrm{max} = t\mrm{H} \delta_{s}(t\mrm{H})^{-3/2}$ , $a^*$ can be evaluated as $a^* \simeq \text{max} \left(a_{(1)}^*, a_{(2)}^*\right)$ where
\begin{equation}
a_{(1)}^*\simeq\frac{2}{5}\sqrt{\frac{3}{5}}q\delta_{s}(t\mrm{H})^{-1/2}\,,~~~~~a_{(2)}^*\simeq\frac{2}{5}\mathcal{I}\sigma\mrm{H} \delta_{s}(t\mrm{H})^{-3/2}\,.
\end{equation}
The fact that a smaller $\delta_{s}(t\mrm{H})$ leads to a larger final value for $a^*$, can be explained by noticing that $t\mrm{max}\propto \delta_{s}(t\mrm{H})^{-3/2}$. Hence for a smaller $\delta_s(t\mrm{H})$, it takes a longer time to reach the nonlinear phase and consequently angular momentum has a longer time to grow.

The finite duration of the early matter-dominated epoch puts a lower bound on $\delta_{s}(t\mrm{H})$. Demanding $t\mrm{max}<t_\text{end}$, where $t_\text{end}$ marks the end of the early matter-dominated era, leads to $\delta_{s}(t\mrm{H})\geq\delta_\text{fd}\equiv(t\mrm{H}/t_\text{end})^{2/3}$ for PBHs formation. The other lower bound on $\delta_{s}(t\mrm{H})$ is set by requiring $a^*\leq1$ or equivalently $\delta_{s}(t\mrm{H})\geq\delta_\text{th(1)}\equiv\frac{3\times 2^2}{5^3}q^2$ and $\delta_{s}(t\mrm{H})\geq\delta_\text{th(2)}\equiv\left(\frac{2}{5}\mathcal{I}\sigma\mrm{H}\right)^{2/3}$. All of these conditions can be summarized as $\delta_{s}(t\mrm{H})\geq \text{max}(\delta_\text{th(1)}, \delta_\text{th(2)}, \delta_\text{fd})$. If $\delta_\text{fd}<\delta_\text{th(2)}$, the effect of finite duration is negligible, otherwise PBHs formation and the probability of formation of PBHs with large spin are severely suppressed.
In this paper we assumed that $\delta_\text{fd}<\delta_\text{th(2)}$. It can be shown that $\delta_\text{th(1)}(q_c)=\delta_\text{th(2)}(q_c)$ where $q_c \equiv \sqrt{2/3}(5/2)^{7/6}\mathcal{I}^{1/3}\sigma\mrm{H}^{1/3}$. 

Since for $\delta_{s}(t\mrm{H})=\sqrt{5/3}\mathcal{I}q^{-1}\sigma\mrm{H}$ 
we have $a_{(1)}^*=a_{(2)}^*$, there is a transition point, $a_{\rm t}^*=(2/5)(3/5)^{3/4}\mathcal{I}^{-1/2}q^{3/2}\sigma\mrm{H}^{-1/2}$, at which the behaviour of $a^*$ is changing
\begin{equation}
a^*\simeq  \left\{
        \begin{array}{ll}
            a_{(2)}^* & \quad \delta_{s}(t\mrm{H}) \leq \sqrt{5/3}\mathcal{I}q^{-1}\sigma\mrm{H}, \\
            a_{(1)}^*& \quad \delta_{s}(t\mrm{H}) \geq \sqrt{5/3}\mathcal{I}q^{-1}\sigma\mrm{H}\,,
        \end{array}
    \right.
\end{equation}
or in terms of $q_c$, $a_{\rm t}^*=(q/q_c)^{3/2}$; a $q>q_c$ leads to $a_{\rm t}^*>1$ which is not acceptable. Since 
\begin{equation}
 \left(a_{(1)}^*\right)^{-2}=\left(\frac{2}{5}\sqrt{\frac{3}{5}}q\right)^{-2}\delta_{s}(t\mrm{H})\,,~~~~~ \left(a_{(2)}^*\right)^{-2/3}=\left(\frac{2}{5}\mathcal{I}\sigma\mrm{H}\right)^{-2/3}\delta_{s}(t\mrm{H})\,,
\end{equation}
$\left(a_{(1)}^*\right)^{-2}$ and $\left(a_{(2)}^*\right)^{-2/3}$ inherit Gaussian distributions
\begin{equation}
\left(a_{(1)}^*\right)^{-2}\sim \mathcal{N}\left[0,\left(\frac{2}{5}\sqrt{\frac{3}{5}}q\right)^{-4}\sigma\mrm{H}^2\right],~~~~~\left(a_{(2)}^*\right)^{-2/3}\sim \mathcal{N}\left[0,\left(\frac{2}{5}\mathcal{I}\right)^{-4/3}\sigma\mrm{H}^{2/3}\right],
\end{equation}
where $\mathcal{N}(\mu,\sigma^2)$ represents a Gaussian distribution with mean $\mu$ and variance $\sigma^2$.

Therefore one can describe the spin distribution of PBHs with the following piecewise distribution
\begin{equation}
n(a^*)= \frac{1}{N}\left\{
        \begin{array}{ll}
            n_\text{(1)}(a^*) \dfrac{n_\text{(2)}(a_{\rm t}^*)}{n_\text{(1)}(a_{\rm t}^*)}& \quad 0 \leq a^* < a_{\rm t}^*, \\
           n_\text{(2)}(a^*) & \quad a_{\rm t}^* \leq a^* \leq 1\,,
        \end{array}
    \right.
\end{equation}
where
\begin{equation}
 n_\text{(1)}(a^*) {\rm d} a^*\propto \frac{1}{a^{*3}}\text{exp}\left(-\frac{1}{2\sigma\mrm{H}^2}\frac{3^2 2^4}{5^6}\frac{q^4}{a^{*4}}\right) {\rm d}a^*\,,~~~~~n_\text{(2)}(a^*) {\rm d}a^*\propto \frac{1}{\left(a^*\right)^{5/3}}\text{exp}\left[-\frac{1}{2\sigma\mrm{H}^{2/3}}\left(\frac{2}{5}\mathcal{I}\right)^{4/3}\frac{1}{\left(a^*\right)^{4/3}}\right]{\rm d}a^*\,,
\end{equation}
and $N$ is the normalization factor.

\subsection{PBH spin distribution from inspirals}
\label{sec:spin_dist_insp}
In the early stages of the evolution of Universe, a sufficiently large ensemble of PBHs may experience mergers if the binary capture rate becomes larger than the expansion rate of the Universe and the inspiral phase ends prior to the Hawking evaporation of PBHs. Ref.~\cite{Arbey:2019zsx} has studied different timescales which are relevant to mergers in a population of PBHs in early Universe. The merger rate could be enhanced if PBHs form in clusters, a hypothesis that will be testable in future experiments looking for
CMB $\mu$-distortion, as proposed recently~\cite{DeLuca:2021hcf}.
If PBHs undergo several mergers before evaporating, the angular momentum gained during each merger causes the spin distribution of PBHs to converge to a universal distribution that is relatively independent of the mass of PBHs, the initial spin distribution of the first generation of PBHs, and the number of merger generations~\cite{Fishbach:2017dwv}. Although Ref.~\cite{Fishbach:2017dwv} considered solar mass black holes, their study is also applicable to PBHs.

The universal hierarchical merger spin distribution in~\cite{Fishbach:2017dwv} has been shown numerically to appear after four merger generations, and to peak at $a^* \simeq 0.7$, with nonzero support over $0.4 \lesssim a^*\lesssim 0.9$. To understand why this universal spin distribution emerges, one needs to consider major contributions to the spin following a merger which consist of the individual spins of the two individual black holes, and the orbital angular momentum of the binary. For equal mass binary black holes the orbital angular momentum dominates over the contribution from the individual spins. Numerical simulations show that merger of non-spinning binary black holes of equal mass will result in a final black hole with $a^*\simeq 0.6864$~\cite{Hofmann:2016yih}.
The spins of the binary black holes can become important and even cancel the orbital angular momentum if they are sufficiently large and anti-aligned to the orbital angular momentum, and the mass ratio needs to be sufficiently small. This is basically why major mergers (with mass ratio $\sim1$) give rise to black holes with high spin  distributions, peaked at $a^* \simeq 0.7$.

A slightly different hierarchical merger spin distribution is reported by~\cite{Doctor:2021qfn} based on the priors from LIGO/VIRGO data for mergers limited to the Milky Way. This distribution also peaks at $a^* \simeq 0.7$.

\acknowledgments
The work of P.S. and B.S. is supported in part by NSF grant ${\rm PHY}$-${\rm 2014075}$. The work of K.S. is supported by DOE Grant desc0009956. 


\end{document}